\documentclass{aa}  

\usepackage{graphicx}
\usepackage{txfonts}
\usepackage{color}
\usepackage{lscape}
\usepackage[colorlinks = true,
            linkcolor = blue,
            urlcolor  = blue,
            citecolor = blue]{hyperref}
\usepackage{soul}
\usepackage{comment}

\newcommand{\lgalaxies}{\textsc{L-Galaxies2020}}

\newcommand{\mstar}{\mbox{$M_{*}$}}
\newcommand{\msun}{\mbox{$M_{\odot}$}}
\newcommand{\mdust}{\mbox{$M_{\rm dust}$}}
\newcommand{\mcold}{\mbox{$M_{\rm cold}$}}
\newcommand{\hi}{{\sc Hi}}        
\newcommand{\htwo}{H$_2$}

\newcommand{\mmtauHI}{$1.97\,\pm\,0.10$}
\newcommand{\altauDust}{$2.26\,\pm\,0.22$}

\newcommand{\tauDust}{$\tau_{d}$}
\newcommand{\tauGas}{$\tau_{HI}$}

\newcommand{\jntauDust}{$2.28\,\pm\,0.1$}
\newcommand{\jntauHI}{$1.92\,\pm\,0.09$}

\newcommand{\sampleLgal}{$2050$}

\begin{document}

   \title{The fate of the interstellar medium in early-type galaxies.}
   \subtitle{IV. The impact of stellar feedback, mergers, and black holes
on the cold interstellar medium in simulated galaxies} 
   \titlerunning{The fate of ISM in ETGs}
   
\author{Jakub Nadolny\inst{1}
\and Michał J. Michałowski\inst{1,2} 
\and Massimiliano Parente\inst{3,4} 
\and Jens Hjorth\inst{5} 
\and Christa Gall\inst{5} 
\and Aleksandra Leśniewska\inst{1,5} 
\and Mart\'in Solar\inst{1} 
\and Przemysław Nowaczyk\inst{1} 
\and Oleh Ryzhov\inst{1}
}

   \institute{Astronomical Observatory Institute, Faculty of Physics, Adam
Mickiewicz University, ul.~S{\l}oneczna 36, 60-286 Pozna{\'n}, Poland 
   \and Institute for Astronomy, University of Edinburgh, Royal Observatory,
Blackford Hill, Edinburgh, EH9 3HJ, UK 
          \and SISSA, Via Bonomea 265, I-34136 Trieste, Italy 
          \and INAF, Osservatorio Astronomico di Trieste, via Tiepolo 11,
I-34131, Trieste, Italy 
        \and DARK, Niels Bohr Institute, University of Copenhagen, Jagtvej
155, DK-2200 Copenhagen N, Denmark 
          \\
        \email{quba.nadolny@gmail.com}
             }
             
   \date{Received ---; accepted ---}

  \abstract%
  {Removing the cold interstellar
medium (ISM) from a galaxy is essential to quenching star formation, however,
the exact mechanism behind this process remains unclear.} 
  {The objective of this work is to find the mechanism responsible for dust
and gas removal in simulated early-type galaxies.}
   {We studied a statistically significant sample of massive (\mstar$\,>\,10^{10}\msun$),
simulated early-type galaxies in a redshift range of 0.02--0.32 in the context
of its ISM properties. In particular, we investigated the cold dust and gas
removal timescales, the cold gas inflows, and their relation with black hole
mass. We also investigated the evolution of galaxies in the {dust mass and star formation rate (SFR)} plane and the influence of merger events. Finally, we broke
down the dust destruction mechanisms to find which (if any) of the implemented
processes dominate as a function of a galaxy's stellar age.}
   {We find a good agreement with previous observational works dealing with
the timescales of dust and \hi\ removal from early-type galaxies. When considering
the dust-to-stellar-mass ratio as a function of time in simulations, we recovered
a similar decline as in the observational sample as a function of stellar
age, validating its use for timing the ISM decline. Moreover, we recovered
the observed relation between dust mass and the SFR for actively star-forming
galaxies, as well as that of passive early-type galaxies. We also show that
starburst galaxies form their own sequence on the dust mass and SFR plot in
the form of $\log(M_{\rm dust, SB})= 0.913\times \log({\rm SFR}) + 6.533,$
with a $2\sigma$ scatter of 0.32. Finally, we find that type II supernova
reverse shocks dominate the dust destruction at the early stages of early-type
galaxy evolution; however, we also see that at later times, stellar feedback
becomes more important. We show that merger events lead to morphological
transformations by increasing the bulge-to-total stellar mass ratio, followed
by an increase in black hole masses. The black hole feedback resulting from
radio mode accretion prevents the hot halo gas from cooling, indirectly leading
to a decrease in the SFR.}
   
   {}


   \maketitle

\section{Introduction}
\label{sec:intro}

Galaxy quenching is the process by which star formation activity decreases
significantly or stops altogether. This cessation of star formation has a
profound impact on the fundamental processes that govern the galaxy evolution.
This is why significant efforts are dedicated to finding the primary mechanism
responsible for quenching, while also considering the possibility that there
could be multiple mechanisms at work.

To stop star formation in a galaxy, it is necessary to either prevent gas
inflows, remove or destroy the cold interstellar medium (ISM), or make the
cold gas unable to form stars. To date, several processes have been proposed.
These mechanisms operate on different time and physical scales \citep{Bell2012ApJ...753..167B,Cheung2012ApJ...760..131C,Hjorth2014ApJ...782L..23H}.
A schematic illustration was presented by \cite{Man2018NatAs...2..695M},
where the authors identified five broad classes of quenching mechanisms:
1) gas does not accrete onto the galaxy; 2) gas does not cool; 3) cold gas
does not form stars; 4) cold gas is rapidly consumed; and 5) gas is removed.

Cold interstellar gas can be consumed by star formation (astration; \citealt{Peng2015Natur.521..192P}),
or it can be ionised either by supernovae (SNe; \citealt{Dekel1986ApJ...303...39D,Muratov2015MNRAS.454.2691M})
or hot low-mass evolved stars \citep{Herpich10.1093/mnras/sty2391}. Additionally,
active galactic nuclei (AGNs) can be responsible for heating and removing
considerable amounts of gas leading to the quenching of star formation \citep{diMatteo2005Natur.433..604D,Piotrowska2022MNRAS.512.1052P}.
It has also been shown that morphological transformation can stop star formation
by merger events and by building up the bulge  \citep{Schawinski2014MNRAS.440..889S}.
This bulge stabilizes gas and prevents it from collapsing, leaving the galaxy
as a bulge-dominated, elliptical, compact early-type galaxies \citep[ETGs;
see mass-size relation, e.g.][]{Mowla2019ApJ...880...57M,Nadolny2021A&A...647A..89N}.
This process is known as morphological quenching \citep{Martig2009ApJ...707..250M,Gensior2020MNRAS.495..199G}.

With respect to dust, several processes governing its production and destruction
have been proposed. Such production processes include: circumstellar shells
around evolved stars such as asymptotic giant branch stars or red giants,
SN remnants, and other non-stellar processes like grain growth in the ISM
\citep{Knapp1992ApJ...399...76K,Jones2004ASPC..309..347J,Matsuura2011Sci...333.1258M,Morgan2003MNRAS.343..427M,Gall2011A&AReview,Nanni2013MNRAS.434.2390N,Michalowski2010A&A...522A..15M,Michalowski2015A&A...577A..80M,Ola_dust_production2019A&A...624L..13L}.
 The destruction processes include astration, SN reverse shock waves, ionising
emission from planetary nebulae (PNs), galactic outflows, or AGNs \citep{Bianchi2005,Clemens2010A&A...518L..50C,Gall2018ApJ...868...62G,Michalowski_dust2019}
acting through sputtering, shattering, and evaporation. See also reviews
by \citet{Jones2004ASPC..309..347J} and \citet{Gall2011A&AReview} and references
therein.

In addition to the mentioned mechanisms acting on the ISM within a galaxy,
an external origin to explain the {\bf evolution} of the ISM has also been proposed
for ETGs. This newly accreted gas (e.g. from mergers) may trigger star formation
with short depletion times \citep[leading to starbursts;][]{Nadolny2023ApJ...952..125N}.
As a consequence, it may then lead to the production and destruction of dust
\citep{vanGorkom1989AJ.....97..708V,Naab2006ApJ...636L..81N,Sanjuan2012A&A...548A...7L}.
Thus, the determination of the ISM removal timescale, $\tau$, is essential
to improving our understanding of the specific process that plays a major
role in the quenching of star formation. 

Due to the destruction, recycling, and replenishing of the gas and dust content
in actively star-forming galaxies \citep{Donevski2020A&A...644A.144D}, it
is challenging to estimate $\tau$ for this galaxy population. As  shown
via high-resolution hydrodynamical simulations \citep{Alvar2022MNRAS.516.2272S},
the emergence of a galactic disc and the dominance of rotational velocity
over dispersion velocity enables bursts of star formation. Therefore, following
\cite{Michalowski_dust2019}, the most appropriate sample to study the ISM
removal timescale is a sample of galaxies with detectable dust or gas, along
with a low star formation rate (SFR). These are characteristics of ETGs that
are mostly found below the main sequence (MS) of star-forming galaxies (SFGs).
 Furthermore, \citet{Magdis2021A&A...647A..33M} showed passive evolution
of dust-to-stellar mass ratio for quiescent galaxies between $z\sim$ 2 and
1, with a steep decrease with decreasing redshift. This decrease is much
steeper as compared to SFG at low-$z$.

In the last few decades, a great effort has been made to tackle the quenching
mechanism in the framework of cosmological simulations. Within the whole
spectrum of different types of simulations, two main branches have emerged.
One of these are simulations based on semi-analytic models \citep[SAMs][]{deLucia2004MNRAS.349.1101D,DeLuciaMillennium2007MNRAS.375....2D,Springel_Milleniu2005Natur.435..629S,Henriques2015}.
The second type are hydrodynamical simulations \citep{Vogelsberger2014MNRAS.444.1518V,Hopkins2014MNRAS.445..581H,Schaye2015MNRAS.446..521S,Pillepich_TNG_description2018,Whitaker_2021,Alvar2022MNRAS.516.2272S,Lorenzon2024arXiv240410568L}.
Each type of simulation focuses on different spatial and temporal scales,
with different mass resolutions. 

In this work, we investigate the atomic hydrogen \hi\ gas and dust removal
timescales using the \lgalaxies\ SAM \citep{HenriquesLGal2020}. In particular,
we adopted the \citet{Parente2023MNRAS.tmp..881P} version of the model, which
includes a detailed treatment of dust evolution within galaxies. { The SAM
parameters have been tuned to reproduce a broad range of galaxy properties
\citep{HenriquesLGal2020,Parente2023MNRAS.tmp..881P}, but not the main quantities
analyzed in this work, namely, the dust (and gas) removal timescales. Therefore,
we employed our SAM to provide a robust physical interpretation of observations,
particularly concerning the quenching mechanism and dust-related processes.}
Our goal of this study is to investigate the ISM behavior in simulated ETGs.
In particular, we study ISM removal timescales and the influence of stellar
mass, BH mass, and redshift on the ISM. Furthermore, we study the evolution
of individual galaxies in the dust mass versus SFR plane. Finally, we investigate
the mechanisms of the destruction and removal of dust in the selected galaxy
samples. 

This article is organised as follows. In Sect. \ref{sec:data}, we present
the observational and simulated data, together with a short description of
the essential characteristics of the \lgalaxies\ models. In Sect. \ref{sec:results},
we present the results of the dust and \hi\ removal timescales for the selected
sample (Sect. \ref{sec:removalTimescale}) for the sample divided in bins
of different parameters (Sect. \ref{sec:ISMRemoval_inBins}) and results for
individual galaxies (Sect. \ref{sec:ISMRemoval_individual}). In Sect. \ref{sec:dust_and_star_formation},
we describe our results in terms of the global relation between dust mass
and SFR, also considering the evolution of individual galaxies. In Sect.
\ref{sec:dust_removal_mechanisms}, we examine several dust destruction and
removal mechanisms. In Sect. \ref{sec:conslusions}, we present our discussion
and conclusions. Throughout this paper, we use a cosmological model with
$h$ = 0.673, $\Omega_\Lambda = 0.685$, and $\Omega_m = 0.315$ \citep{Planck2014}.
The initial
mass function (IMF) from  \citet{Chabrier2003PASP..115..763C} was assumed
in the applied SAM.

\section{Data}
\label{sec:data}

\begin{figure*}[ht!]
\includegraphics[width=0.45\textwidth,clip]{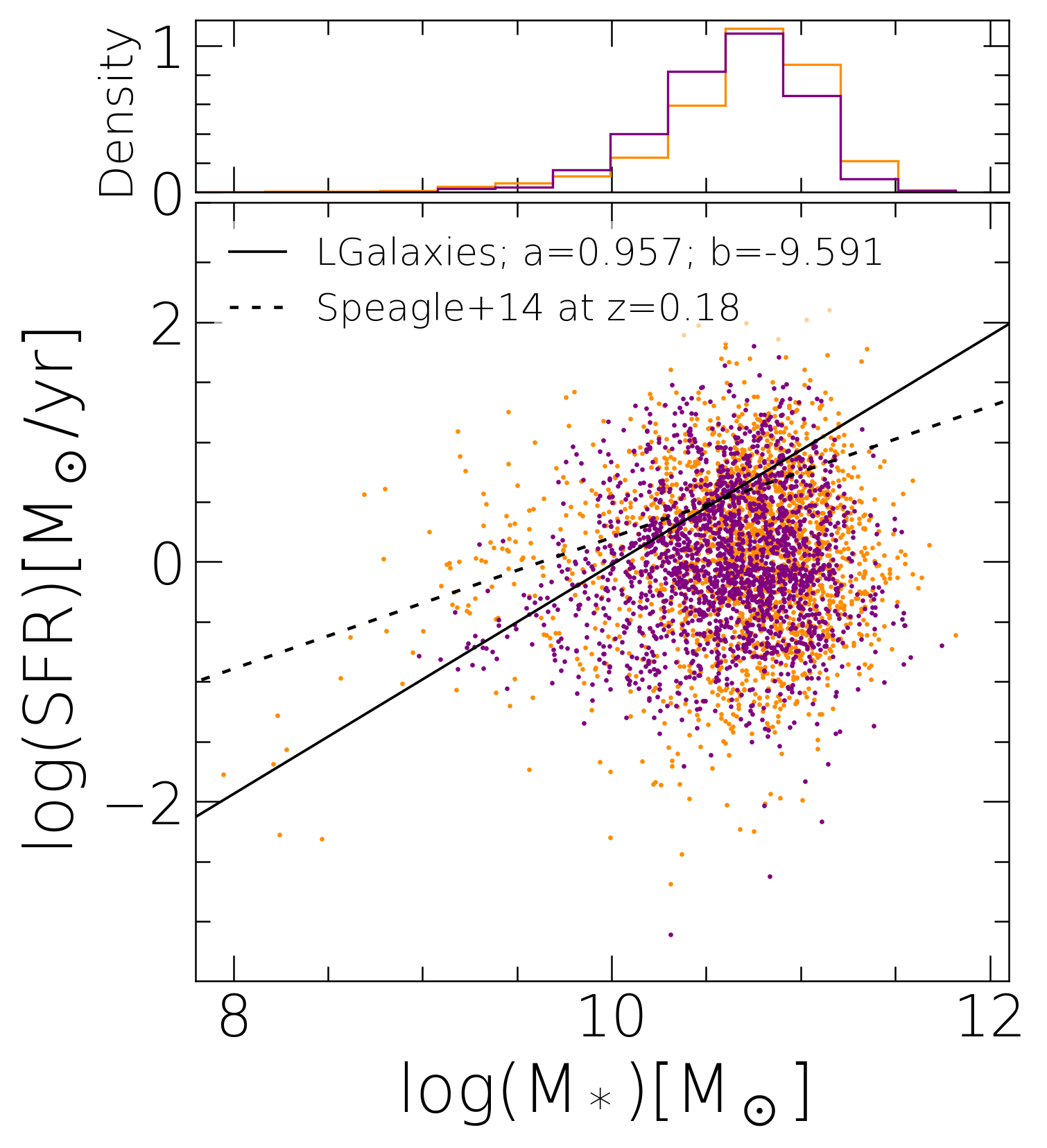}
\includegraphics[width=0.45\textwidth,clip]{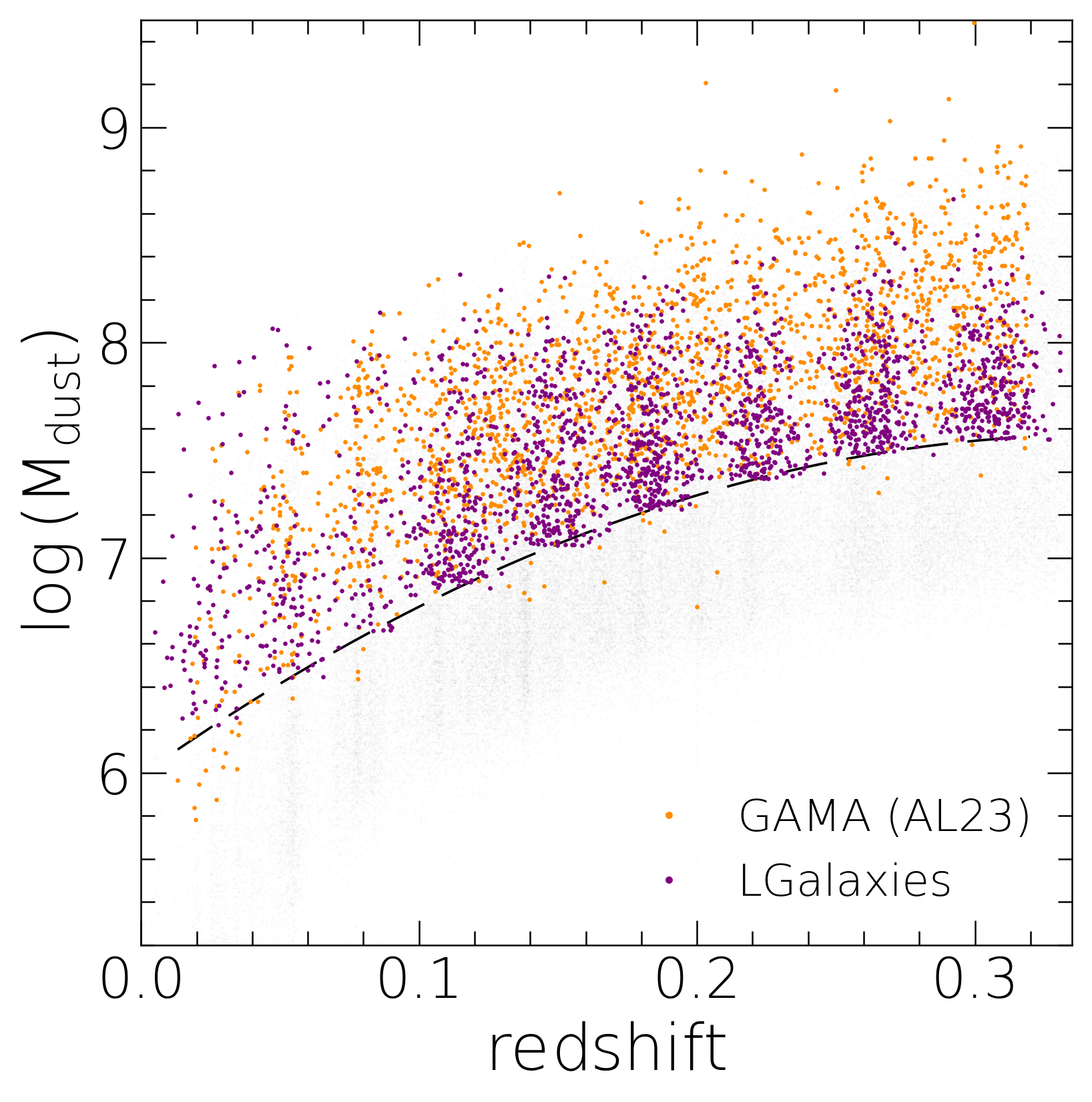}
\caption{SFR as a function of stellar mass (left). The solid line shows the
MS from the \lgalaxies\ simulations estimated using LTGs (see main text for
details) and is a power-law fit described as $\log(SFR) = a \times \log(M_*)
+ b$ with coefficients given in the legend. The dotted line shows the MS
estimated using a prescription from \citet{Speagle2014ApJS..214...15S}, both
at $z=0.18$. Orange and gray points show the GAMA, and \lgalaxies\ ETG samples.
Normalised stellar mass distribution for the mentioned samples (top). Dust
mass as a function of redshift (right). The dashed line shows the dust mass
limit estimated for the GAMA sample. More details are given in Sect. \ref{sec:data_simulations_lgalaxies}.
\label{fig1:MS_dustLim}}
\end{figure*}

\subsection{Observational data}\label{sec:data_observational}
In this work, we made use of samples defined in \cite{Lesniewska2022} and
\cite{Michalowski2022}. In particular, \cite{Lesniewska2022} studied the dust-to-stellar
mass ratio as a function of stellar age of observed ETGs (hereafter the GAMA
sample), whereas \cite{Michalowski2022} studied the \hi-to-stellar mass ratio
as a function of stellar age. We refer to these works for details on the
data used. In short, both samples were selected based on similar criteria:
(i) early-type morpologies were selected using a S\'ersic index \citep{Sersic1963}
with values higher than four ($n > 4$) in \cite{Lesniewska2022} or visual
early-type morphology in \cite{Michalowski2022}; 
(ii) a redshift range between 0.02 and 0.32 was selected to ensure that at
higher redshifts the S\'ersic index selection does not include bulge-dominated
or compact late-type galaxies (LTGs); and (iii) for the dust 
detection, we followed both \citet{Lesniewska2022} and \citet{Michalowski2022},
where a signal-to-noise (S/N) threshold at 250$\mu$m of 3 was used. Due to
the size of the GAMA sample, it was used as a reference for our \lgalaxies\
sample selection. 

\subsection{Simulations}
\label{sec:data_simulations}
In this work, we make use of the predictions of the \lgalaxies\ SAM, which
is the latest version of the Munich semi-analytic model for galaxy evolution
\citep{Henriques2015,HenriquesLGal2020}. In particular, we adopted the version
introduced by \cite{Parente2023MNRAS.tmp..881P}, which accounts for a detailed
model for dust production and evolution, as well as for an updated treatment
of disc instabilities. In this section, we summarise the main features of
this SAM, while we refer to the supplementary material of \cite{HenriquesLGal2020}
for a detailed description of all these processes.

The \lgalaxies\ model starts from the dark matter (DM) haloes merger trees
of the \textsc{Millennium} simulation \citep{Springel_Milleniu2005Natur.435..629S}
and populates them with baryons according to a number of physical processes,
which are important for galaxy evolution. Among them are gas infall into
DM haloes, gas cooling, disc and bulge formation, star formation, merger-driven
starbursts, chemical enrichment, energetic feedback from stars and supermassive
black holes (SMBHs), and environmental processes such as ram pressure and
tidal stripping.

The ISM is represented by a disc which results from the (metallicity and
temperature dependent) cooling of the hot halo gas. Such a disc is spatially
resolved: gas properties are tracked for each simulated object in $12$ concentric
rings. For each of them, the SAM determines the molecular fraction of the
ISM according to the \cite{Krumholz2009ApJ...693..216K} model. Subsequently,
this \htwo\ mass fuels the star formation \citep{Gao2004ApJS..152...63G,Wong2002ApJ...569..157W}.
The star formation law implemented in the \lgalaxies\ is based on the \htwo\
surface density \citep{Bigiel2011ApJ...730L..13B,Leroy2013AJ....146...19L}.
An inverse dependency of star formation on the dynamical time \citep[Sect.~2.2.4
in][]{HenriquesLGal2020} has also been  implemented. This ensures that star
formation is more efficient for galaxies with shorter dynamical times, in
particular at earlier epochs \citep{Scoville2017ApJ...837..150S}. The stars
produce energetic and chemical feedback, which was modelled according to
\cite{Yates2013MNRAS.435.3500Y}. This model traces the abundances of eleven
chemical elements (H, He, C, N, O, Ne, Mg,Si, S, Ca, and Fe) produced in
supernovae (SNe type Ia and II) and in asymptotic giant branch (AGB) stars
released by their winds. Both the feedback and chemical enrichment are modelled
separately for each ring of the disc. Tracing the evolution of baryons in
each ring is essential for the evaluation of the chemical evolution, \htwo\
fraction, and SFRs. For this work, however, we use integrated properties
(e.g. total stellar mass, total SFR, and total dust mass) of the selected
sample -- and not the values evaluated for each ring.

In addition, star formation is sensitive to merger events that may trigger
a burst of star formation. As for the descendant galaxy, the SAM models different
outcomes for minor and major mergers. In a major merger, the discs of the
progenitors are destroyed, and all the stars form the bulge of the resulting
(descendent) galaxy. Any star that forms during this event is added to the
bulge, increasing its mass. During a minor merger, the disc of a larger progenitor
survives and accretes the cold gas from the smaller galaxy. All the stars
from the smaller galaxy are included in the bulge, while newly formed stars
are added to the disc of the descendant galaxy. Thus both minor and major
mergers shape the morphology of a galaxy, in particular by increasing the
bulge mass. 

Disc instabilities can also result in the bulge growth, although they mostly
affect the morphology of intermediate-mass galaxies. The modelling of this
process has been revised in the SAM version we adopt \citep{Parente2023MNRAS.tmp..881P},
leading to a better agreement with the observed abundance of local galaxies
with different morphologies.

Merger events and disc instabilities\footnote{This in-situ SMBH growth channel
is not present in the \cite{HenriquesLGal2020} SAM, but it has been included
in the \cite{Parente2023MNRAS.tmp..881P} version. Recently, \cite{Parente24}
have shown the relevance of this channel, within the context of this SAM.}
also drive the growth of the central SMBHs. In the \lgalaxies\ the growth
of a BH is through radio or quasar mode. While the quasar mode is mostly
responsible for the black hole growth, in the adopted SAM it produces no
feedback. On the other hand, the radio mode accretion of the hot gas provides
feedback that increases with increasing black hole mass and hot gas mass
(Sect. 1.14.2 of the supplementary material of \citealt{HenriquesLGal2020}).
The radio mode feedback is crucial since it leads to the quenching of star
formation by preventing hot gas from cooling \citep{Henriques2017MNRAS.469.2626H}.

Finally, the SAM self-consistently models the formation and evolution of
dust grains within galaxies. The dust model was introduced in \citet[with
details given in their Sect. 2.1]{Parente2023MNRAS.tmp..881P}. It was inspired
by state-of-the-art models already successfully adopted in hydrodynamical
simulations \citep{Granato21,Parente22}. The model is based on the two-size
approximation \citep{Hirashita2015MNRAS.447.2937H} and it follows the evolution
of two sizes and two chemical compositions of grains (i.e. silicate and carbonaceous,
MgFeSiO$_4$ and C, with radii of $0.05$ and $0.005 \, \mu{\rm m}$). In short,
large grains are assumed to be produced by AGB stars and type II SNe and
are then ejected into the cold and hot gas of galaxies. Different processes
that affect the evolution of grains are modelled in the SAM. These include
grain growth by accretion of gas-phase metals, destruction by (type Ia and
II) SN explosions with a part of the energy released used for heating of
the gas which is then transferred from the cold phase to the hot atmosphere
together with corresponding dust (see \citealt{Parente2023MNRAS.tmp..881P}
and supplementary material in \citealt{HenriquesLGal2020} for details).
Furthermore sputtering in the hot gas, as well as shattering (coagulation)
of large (small) grains is also implemented. The modelling of these processes
is done on the fly and through physically motivated recipes which depend
on the physical properties of the gas phase of simulated galaxies. This model
has been shown to reproduce several scaling relations involving dust at $z<2.5$
and (notably in the context of this work) the dust abundance in LTGs and
ETGs at $z=0$.

\subsection{Sample selection}
\label{sec:data_simulations_lgalaxies}

The selection process of the simulated sample of ETGs has been designed to
 closely follow the selection of galaxies studied in the observational works
to which we are comparing our results. Our selection was undertaken based
on the following criteria: (i) redshift range between 0.02 and 0.32; (ii)
bulge-to-total stellar mass ratio above 0.7; and (iii) dust mass above the
dust mass limit estimated using the GAMA sample to mimic their 250$\mu$m
S/N limit. This dust mass limit is estimated using a polynomial fit to the
GAMA sample dust mass as a function of redshift (Fig. \ref{fig1:MS_dustLim},
dashed line on the right panel) and shifted $2\sigma$ downwards; it
is given as: \[log(M_{\rm dust}) = -13.21 \times z^2 + 9.14\times z + 5.99,\]
where $z$ is redshift. Finally, we have (iv) a random selection of galaxies
in eight redshift bins (with a width of 0.15) to follow the redshift distribution
of the observed GAMA sample.

After this selection process, our ETG \lgalaxies\ sample contains \sampleLgal\
galaxies. This is the exact number of galaxies as presented in \citet{Lesniewska2022}
with very similar physical properties that allow precise comparison. We show
the selected galaxies in Fig. \ref{fig1:MS_dustLim}. 
We note that the stellar masses and dust masses of the selected ETGs are
slightly lower (on average) by 0.12 dex and 0.2 dex, respectively, as compared
to \citet{Lesniewska2022}. In particular, for the same stellar mass, the
simulated galaxies have, on average, 0.3 dex lower dust mass. 
 
In the left panel in Fig. \ref{fig1:MS_dustLim}, we compare the MS estimated
using a prescription from \citet{Speagle2014ApJS..214...15S} at $z = 0.18$,
namely, the same as used in \citet{Lesniewska2022}, and the MS estimated
from simulated LTGs. These simulated LTGs were selected by imposing a bulge-to-total
mass ratio below 0.3 and $z = 0.18$. {Using a standard 0.2 dex scatter we
find that 68\% of our ETGs are found below the MS, consistently with \citet{Lesniewska2022}.
We  describe this particular selection in Sect. \ref{sec:dust_and_star_formation}
in the context of the \citet{daCuhna2010} relation.}

\begin{figure}[t!]
\includegraphics[width=0.48\textwidth,clip]{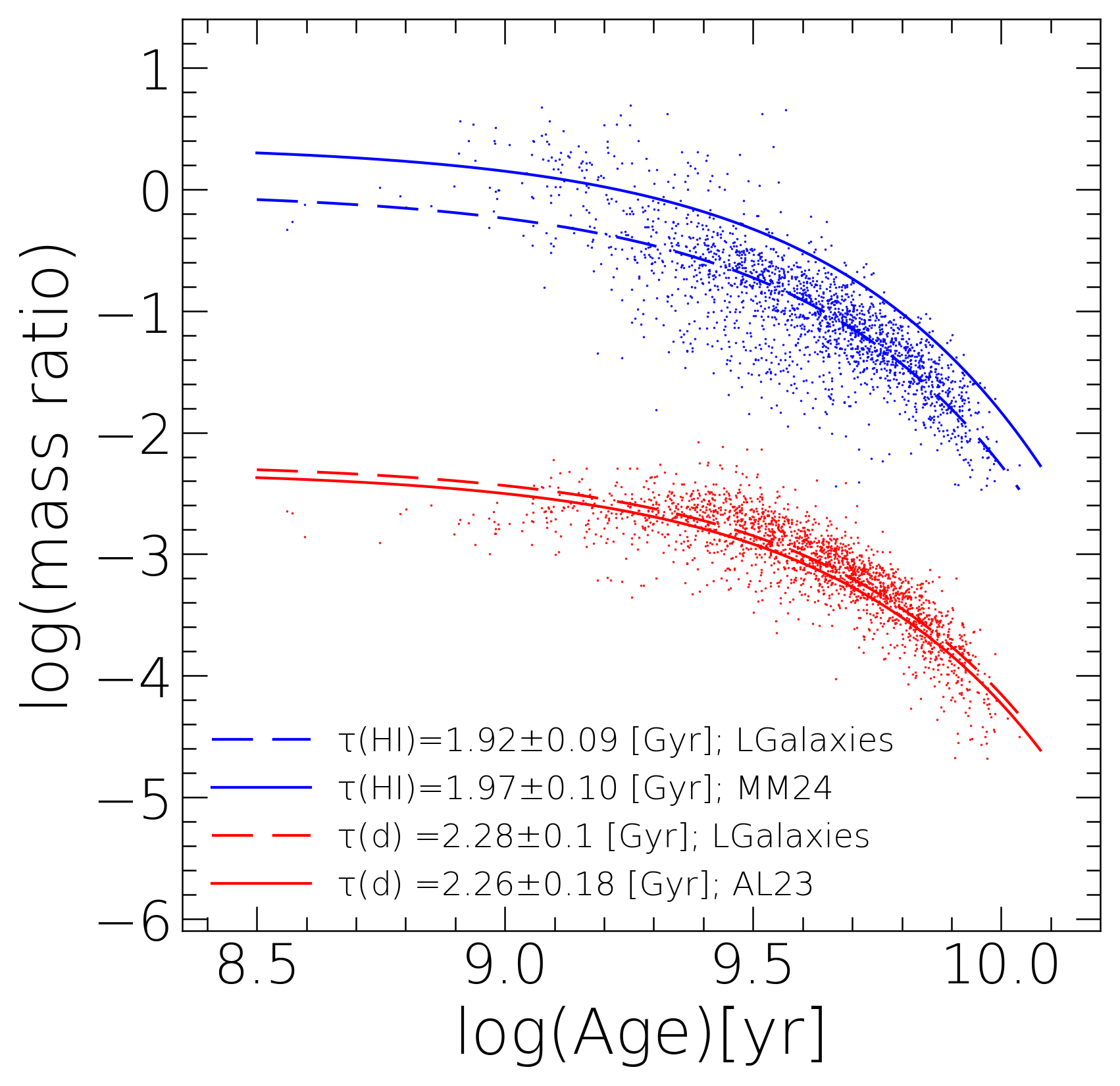}
\caption{\hi-to-stellar and dust-to-stellar mass ratios as a function of
light-weighted stellar age. Exponential fits to \lgalaxies\ ETG samples (dashed
lines), together with observationally derived fits from \citet{Michalowski2022,Lesniewska2022}
(continuous and dotted lines) are shown.
\label{fig2:exponentialFits}}
\end{figure}

\section{Results}\label{sec:results}

\subsection{ISM removal timescales}\label{sec:removalTimescale}

The dust and gas removal timescales are determined using an exponential function
in the form \citep{Michalowski_dust2019} of:
\begin{equation}\label{eq:expo}
  \frac{Mass}{M_*} = Ae^{-{\rm age}/\tau},  
\end{equation}
where {\it Mass} stands for dust or \hi\ gas masses, $A$ is a normalisation
constant, and $\tau$ is the ISM removal timescale. We fit this exponential
function to the dust-to-stellar and \hi-to-stellar mass ratios and obtained
the results shown in Fig. \ref{fig2:exponentialFits},  together with the
observational results. We found dust (\tauDust) and \hi\ (\tauGas) removal
timescales of \jntauDust\ Gyr and \jntauHI\ Gyr, respectively. Errors given
in this work have been estimated as a standard deviation of resulting $\tau$
from 1000 Monte Carlo runs, where we perturbed the mass ratio for all galaxies
independently by a randomly selected value from a Gaussian distribution centered
at $\mu = 0$ and with a width defined as a standard deviation of a distance
from the best-fit line.  
The resulting dust and \hi\ removal timescales are in excellent agreement
with the results from \citet{Lesniewska2022}, where they found \tauDust=\altauDust\
Gyr, and \tauGas=\mmtauHI\ from \citet{Michalowski2022}. Our analysis of
the selected simulated galaxies using the \lgalaxies\ SAM is the first (to
our knowledge) effort to estimate the dust and \hi\ dust removal timescales
in the same way as was done for the observed samples. {Similar result was found in hydrodynamical simulations in very recent work by \citet{Lorenzon2024arXiv240410568L}.}

\subsection{ISM removal timescales and galaxy properties}
\label{sec:ISMRemoval_inBins}
\begin{figure*}[ht!]
\includegraphics[width=0.95\textwidth,clip]{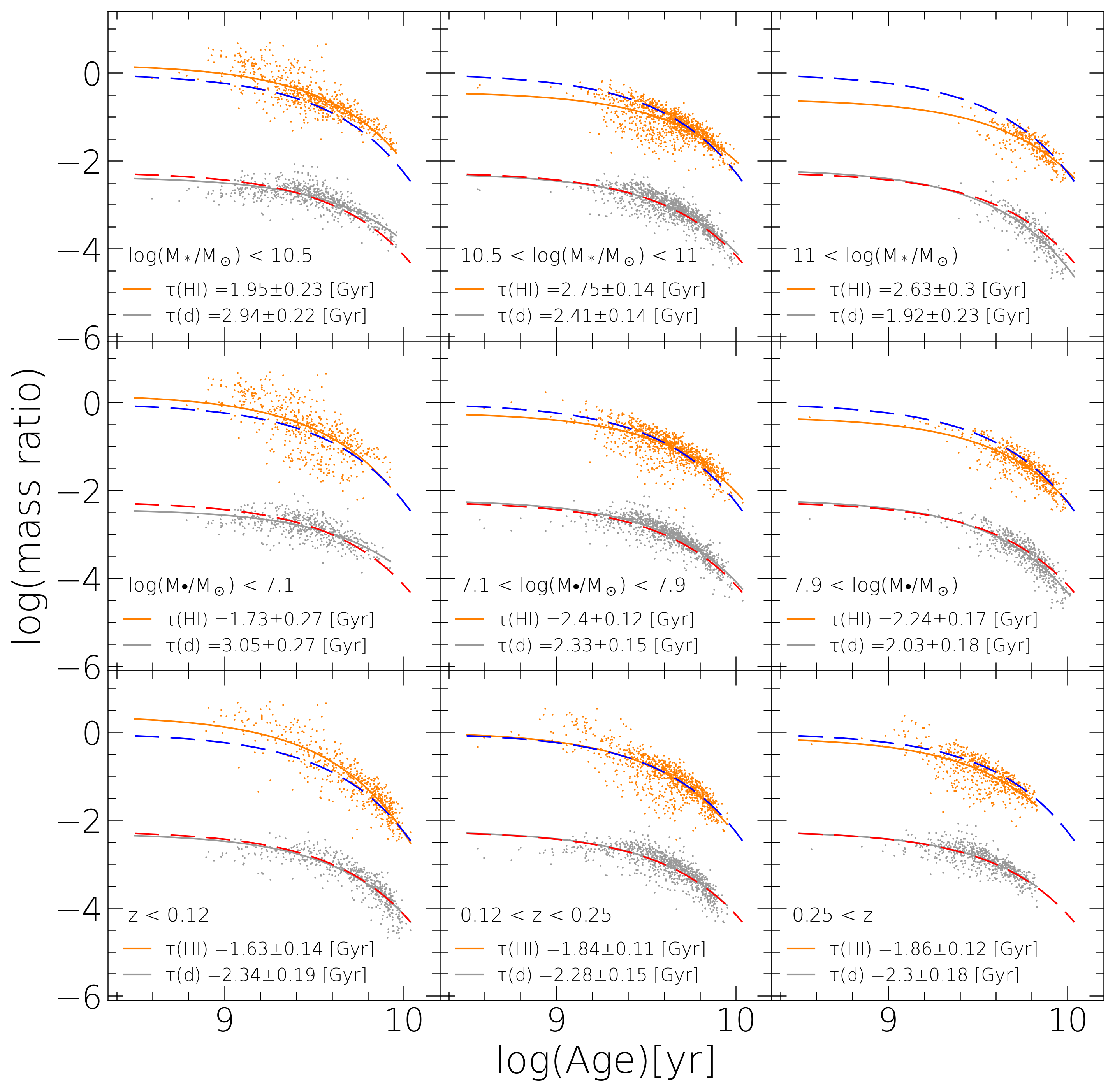}
\caption{\hi-to-stellar and dust-to-stellar mass ratio as a function of light-weighted
age. Orange and grey lines show fits to \hi- and dust-to-stellar mass ratios
for sub-samples divided into stellar mass (top row), the 
black hole-to-stellar mass ratio
(middle), and redshift (bottom row) bins. Blue and red dashed lines show
fits to the overall ETG sample from Fig. \ref{fig2:exponentialFits} and are
given for reference.
\label{fig3:exponentialFits_inBins}}
\end{figure*}

\begin{figure*}[ht!]
\includegraphics[width=0.43\textwidth,clip]{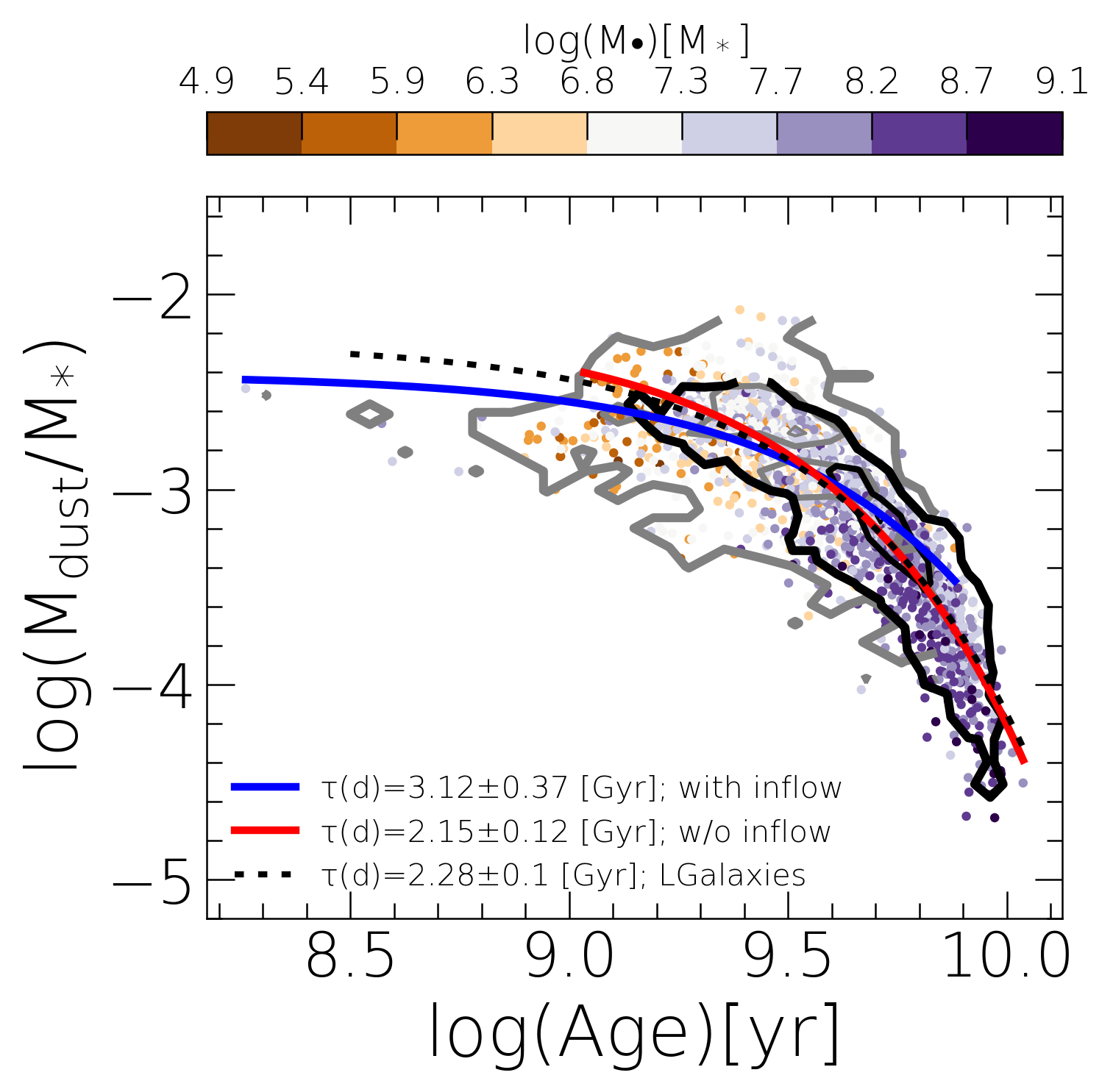}
\includegraphics[width=0.43\textwidth,clip]{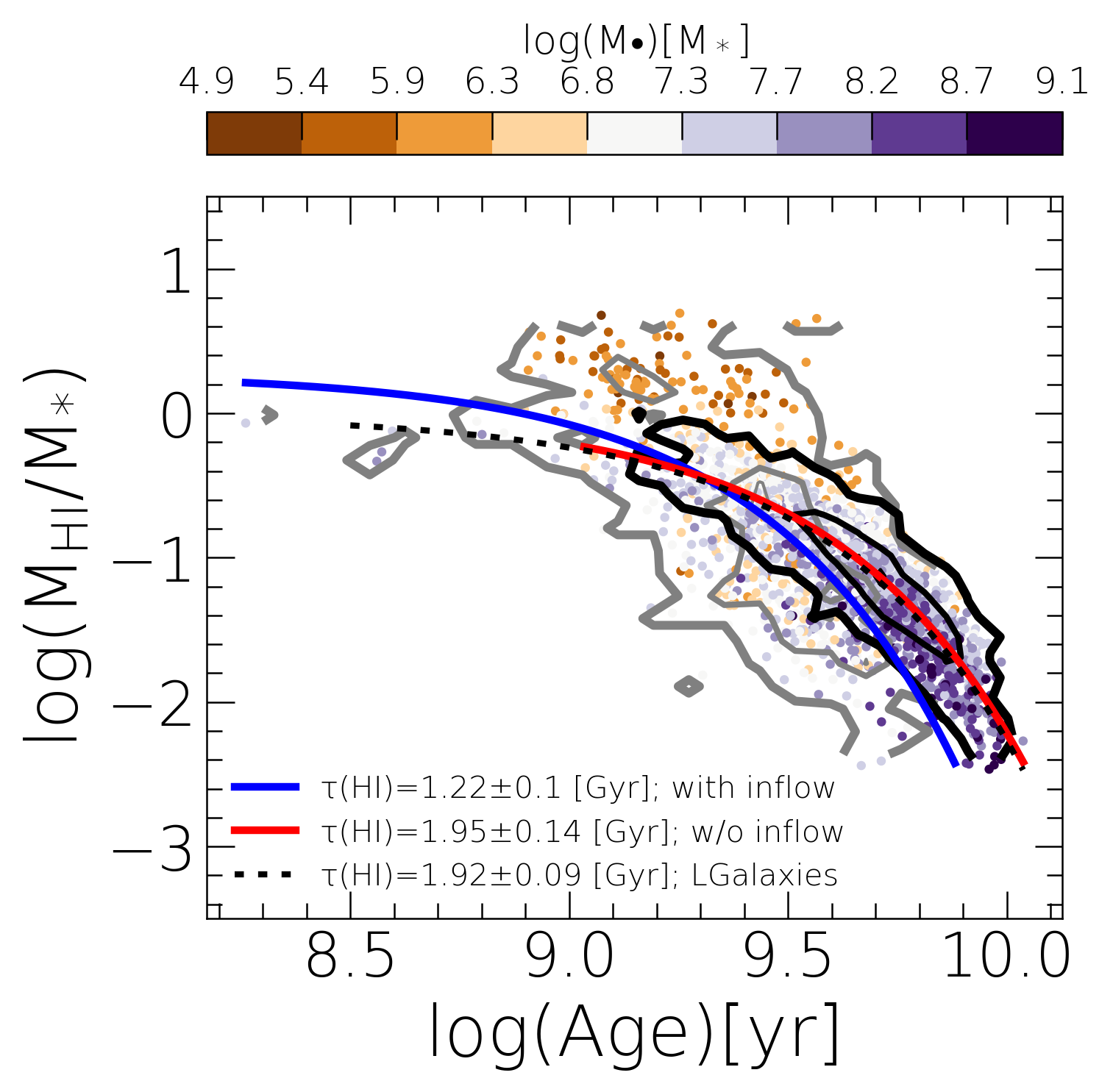}

\includegraphics[width=0.44\textwidth,clip]{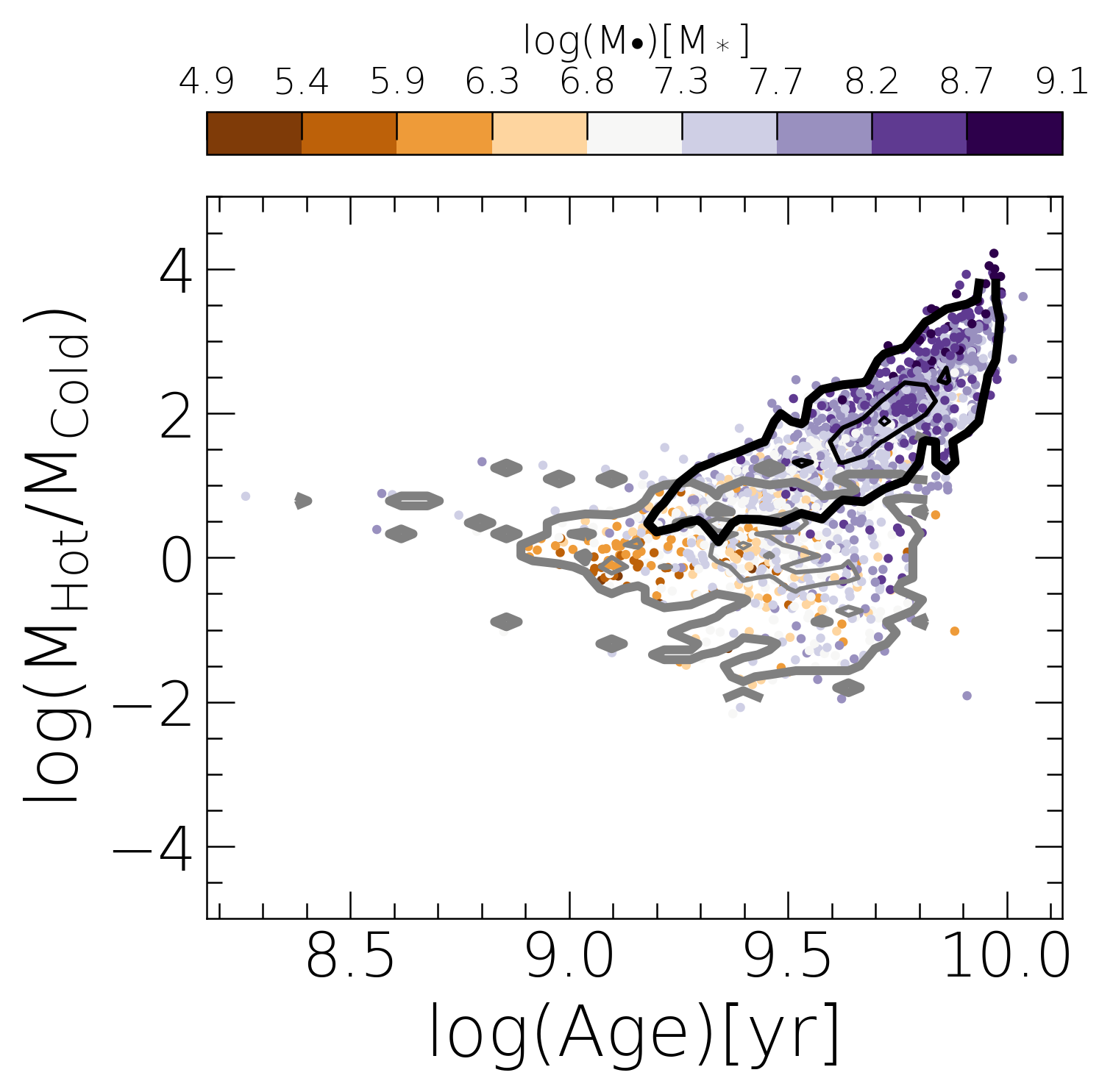}
\includegraphics[width=0.44\textwidth,clip]{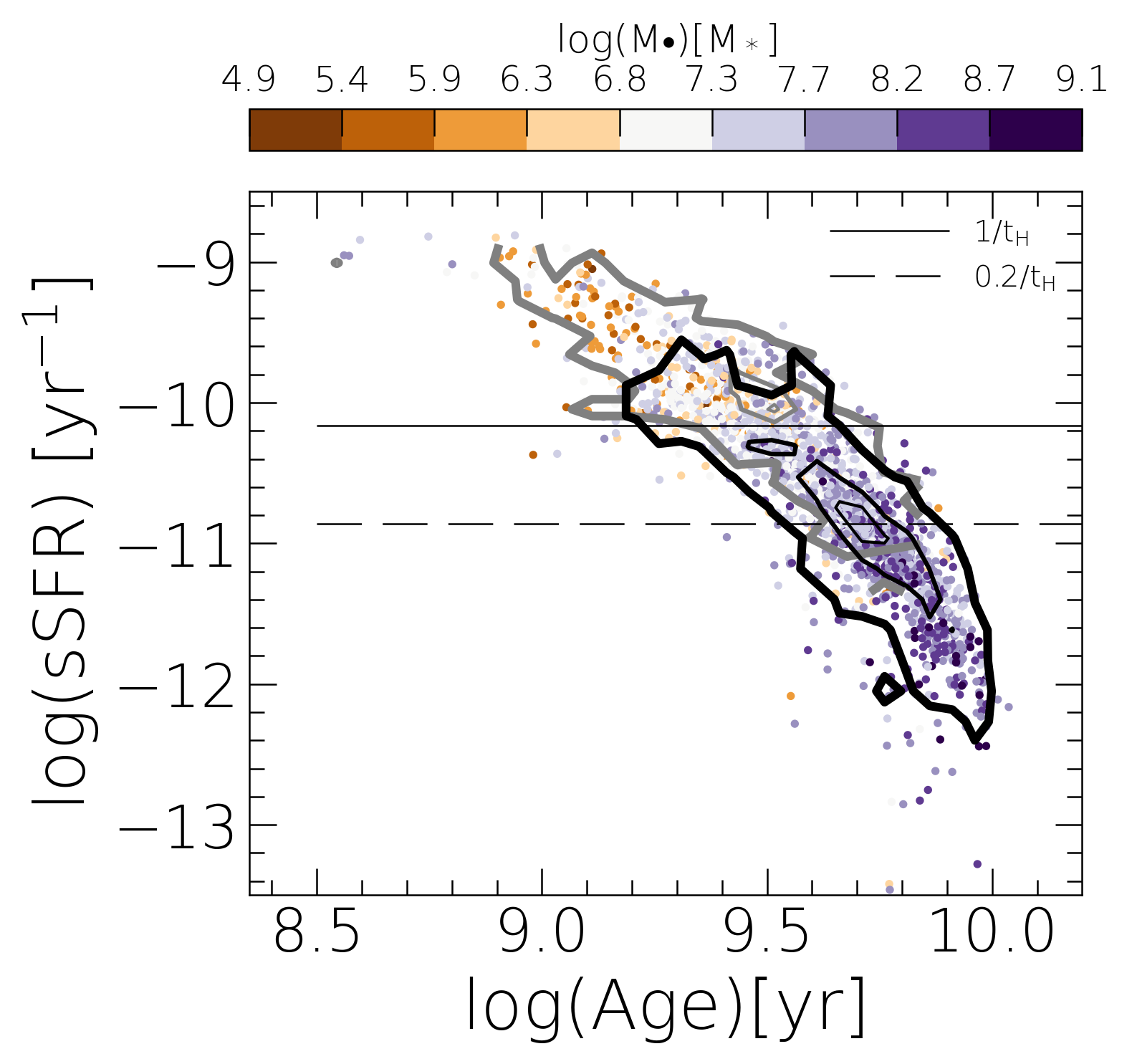}
\caption{Dust-to-stellar mass ratio, and \hi-to-stellar mass ratio as a
function of light-weighted age (top). Blue and red solid curves show fit
to sub-samples with and without the inflow, respectively. The black dotted
line is the fit to the ETGs, the same as in Fig. \ref{fig2:exponentialFits}.
 Hot-to-cold gas mass ratio as a function of age (bottom-left). 
Specific SFR as a function of age (bottom-right). Solid and dashed lines
show $t^{-1}_{H}$ and $0.2\times t^{-1}_{H}$, where $t_{H}$ is the Hubble
time at $z=0$ \citep{Montero_SIMBA_2019MNRAS.490.2139R}. In all panels: points
show our ETG sample colour-coded by the black hole mass, grey and black contours
represent galaxies with and without an inflow of the material between the
current (i.e. the snapshot at which the galaxy was selected) and the previous
snapshot. 
\label{fig:dust-to-stellar_with_BHmass}}
\end{figure*}

In this section, we test whether other galaxy properties affect the ISM removal.
In particular, we investigate the mass ratio dependence on stellar mass,
black hole (BH) mass, and redshift (see Fig. \ref{fig3:exponentialFits_inBins}).
The dust removal timescales vary between 2.03 and 3.05 Gyr; however, most
of these are consistent within the errors and the normalisation of the dust-to-stellar
mass ratio remains roughly constant in all cases (as compared to the fit
to our full ETG sample, shown as a red dashed line).  

In the case of \hi, the variation in the removal timescale and the normalisation
is visible, in particular when using stellar and BH mass bins. The \tauGas\
difference, however, is not statistically significant. For the stellar (BH)
mass bins, the \hi-to-stellar mass ratio normalisation constants, $A,$ are
$0.20\pm 0.12$ ($0.18\pm 0.13$), $-0.42\pm 0.05$ ($-0.22\pm 0.04$), and $-0.58\pm
0.12$ ($-0.31\pm 0.1$) from the low to high stellar (BH) mass. 
The normalisation for the entire sample for \hi-to-stellar mass ratio fit
is $A = 0.04\pm0.06$, while the same normalisation constant from \citet{Michalowski2022}
is $A = 0.37\pm0.05$.
These differences in the normalisation constant $A$ in stellar mass bins
reflect the correlation of the gas-to-dust mass ratio with metallicity, as
shown in \citet{RemyRuyer2014A&A...563A..31R}; namely, it is evident that
more massive (and more metal-rich as a result of the well-known mass-metallicity
relation; e.g. \citealt{Maritza_FP_2010A&A...521L..53L,Nadolny2020A&A...636A..84N})
galaxies have lower gas-to-dust ratios. Moreover, this relation was successfully
reproduced (using the SAM employed in this work) in \citet{Parente2023MNRAS.tmp..881P}.
To test that, we analyze our sample divided in metallicity bins finding
similar normalisation constants of $0.28\pm0.19$, $-0.20\pm0.05$, and $-0.80\pm0.06$
for increasing metallicities. These fits are not shown due to great similarity
with the results using stellar mass bins. 

Considering the dust and \hi\ mass in redshift bins we found no evidence
for evolution with cosmic time (at least up to $z = 0.32$, the redshift limit
studied in this work). Similar results have been found in \citet{Lesniewska2022}.

As shown in the framework of the applied SAM \citep{Henriques2015,HenriquesLGal2020,Parente2023MNRAS.tmp..881P},
the BH mass is correlated with bulge mass consistently with the well-known
relation between the BH mass and stellar dispersion velocity \citep{Magorrian1998AJ....115.2285M,Gebhardt2000ApJ...539L..13G,McConnell2013ApJ...764..184M}.
This correlation is the effect of the interplay between the implemented mechanisms
that control the growth of both, namely, merger events, disc instabilities,
and AGN radio and quasar feedback modes \citep[see Sect. 2.1 in][and the
summary in Sect. \ref{sec:data_simulations}]{Henriques2017MNRAS.469.2626H}.
In particular, the feedback from the radio mode accretion releases energy
that is proportional to the BH mass and hot gas mass. Thus, this energy input
prevents the hot halo gas from cooling down and effectively cuts off the
fuel input for future star formation.

We highlight the importance of the BH effect on simulated galaxies in Fig.
\ref{fig:dust-to-stellar_with_BHmass}, where the dust- and \hi-to-stellar
mass ratio (top panels), hot-to-cold gas ratio, and specific SFR (sSFR; bottom
panels) are shown as a function of age. The (colour-coded) black hole mass
increases with decreasing dust and \hi\ masses and increasing hot-to-cold
gas mass ratio.

In Fig. \ref{fig:dust-to-stellar_with_BHmass}, we also show blue (red) contours
to denote galaxies with (without) an inflow between this and the previous
snapshot. These constitute a total of 18\% (82\%) of our ETG sample. 
The AGN radio-mode feedback is responsible for preventing the hot halo gas
from cooling down and inflowing into the galaxy.  
We estimated the dust and \hi\ removal timescales for both sub-samples. We
find that galaxies with inflow have longer (shorter) dust (\hi) removal timescales
as compared to the sub-sample without inflows (although consistent within
the errors). 
Galaxies with inflow are younger with more dust and cold gas [mean  log(age/yr)=9.4,
log(\mdust/\msun)=7.43, and 
log(\mcold/\msun)=9.7] than galaxies without inflow [mean log(age/yr)=9.7,
log(\mdust/\msun)=7.35, and log(\mcold/\msun)=9.5]. The sub-sample without
inflows shows removal timescales that are in agreement with the timescales
estimated for the full sample (for both mass ratios). Thus, the inclusion
of the galaxies with inflow does not alter the overall results on dust and
gas removal timescales. The bottom-left panel shows the clearest separation
between these two sub-samples. Galaxies with inflow show a larger content
of cold gas, while galaxies without inflow have much more hot gas. To further
investigate this issue, in the bottom-right panel of Fig. \ref{fig:dust-to-stellar_with_BHmass},
we show sSFR as a function of age with the contours representing the same
sub-samples as in the rest of the panels. Galaxies without inflow and higher
BH masses (colour-coded) are found mostly below the quenched galaxy sSFR
limit of $0.2\times t^{-1}_{\rm H}$ (or below the MS), while galaxies with
inflow are above the SFG limit of $t^{-1}_{\rm H}$ \citep{Montero_SIMBA_2019MNRAS.490.2139R}
(or on the MS), where $t_{\rm H}$ is the Hubble time.

\subsection{ISM removal timescales in individual galaxies}
\label{sec:ISMRemoval_individual}

\begin{figure}[t!]
\includegraphics[width=0.48\textwidth,clip]{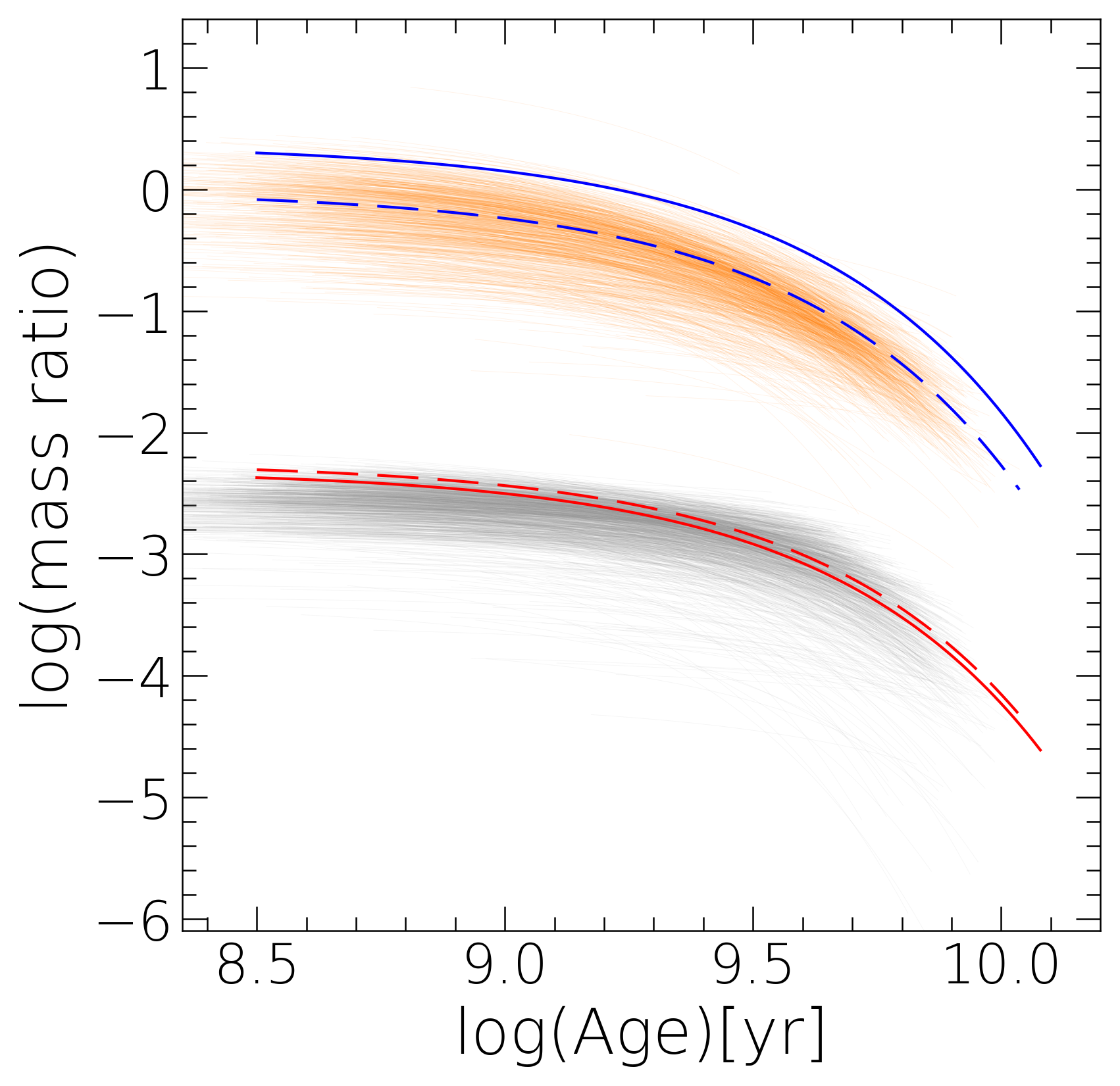}
\caption{\hi-to-stellar and dust-to-stellar mass ratios as a function of
light-weighted stellar age. { Thin lines show exponential fits to our individual
ETG for our sample and their progenitors from which we select fits that have
$\tau$/$\tau_{err} > 1$  
and these constitute 63\% and 47\% of the total for the dust- and \hi-to-stellar
ratios, respectively.} Blue and red lines are the same as in Fig. \ref{fig2:exponentialFits}.
\label{fig2:exponentialFits_individual}}
\end{figure}

By taking advantage of the simulations, it is possible to trace the selected
galaxies backwards in time. This allows us to trace the evolution of an individual
galaxy and its progenitors. Figure \ref{fig2:exponentialFits_individual}
shows the exponential fits to the ISM-to-stellar ratio versus~age (described
in Sect. \ref{sec:removalTimescale}) together with selected fits to the ISM-to-stellar
ratio versus~age for individual galaxies and their progenitors up to $z=1.14$.
To find out whether these individual fits are similar to the fit for the
ETG sample, we estimated the percentage of fits that have relative errors
of the timescale $\tau$ lower than 100\%. In this way, we were able to select
a total of { 63\%} (dust-to-stellar) and { 47\%} (\hi-to-stellar mass ratio)
of the total ETG sample. The fits failing to meet this criterion correspond
to objects with high ISM variations (e.g. due to mergers, see below). These
fits are mostly either shallower (i.e. with much larger, unrealistic $\tau$)
or significantly steeper (i.e. with much smaller $\tau$). 

It is important to point out that the specific characteristic parameters
of the progenitors can significantly differ from the selected early-type
descendent galaxy. This is especially true before a merger or disc instability
occurs during the evolution of a particular galaxy. In other words, the
progenitors of our galaxies might not be early-type galaxies after all (as identified
through the selection process described in Sect. \ref{sec:data_simulations_lgalaxies}).
However, we find that for $\sim$47\,--\,63\% of the sample this does not
have an impact on the fitting process and we can describe the dust-to-stellar
and \hi-to-stellar mass ratio evolution with our method at least down to
light-weighted age of $10^{8.5}$\,yr.

\subsection{Dust and star formation}
\label{sec:dust_and_star_formation}

A strong correlation between dust content and star formation for MS galaxies
was found more than a decade ago \citep{daCuhna2010}. A similar, albeit
with a lower slope, relation for elliptical galaxies  below the MS was first
found by \citet{Hjorth2014ApJ...782L..23H} and this was recently confirmed
using a larger sample in \citet{Lesniewska2022}. In particular, \citet{Hjorth2014ApJ...782L..23H}
analyzed analytical models of dust production and star formation and interpreted
the positions of particular populations of galaxies as evolutionary stages
(starburst, star formation, and quenched star formation). They suggested
that the initial phase is starburst-driven as a result of an increased supply
of gas for star formation (e.g. by a merger). In this phase, while the SFR
is high, the dust mass is built up quickly. After this phase, a normal
star formation takes place which is characterised by a joint decay in dust
mass and SFR resulting in the \citet{daCuhna2010} relation. This particular
evolutionary phase is described with a slope of $\sim1$ (found in this work,
see below, or 0.9 in \citealt{Hjorth2014ApJ...782L..23H}). This phase is
related to the global Schmidt-Kennicutt relation; however, this relation
has a larger slope: between 1.2--1.5 \citep{Schmidt1959ApJ...129..243S,Kennicutt1998ApJ...498..541K}.
The last phase is the quenching of star formation in which a rapid decline
in SFR is observed, while the dust decline is slower. Taking advantage of
the present simulations, here we show how individual galaxies move around
this diagram.

\begin{figure}[t!]
\includegraphics[width=0.48\textwidth,clip]{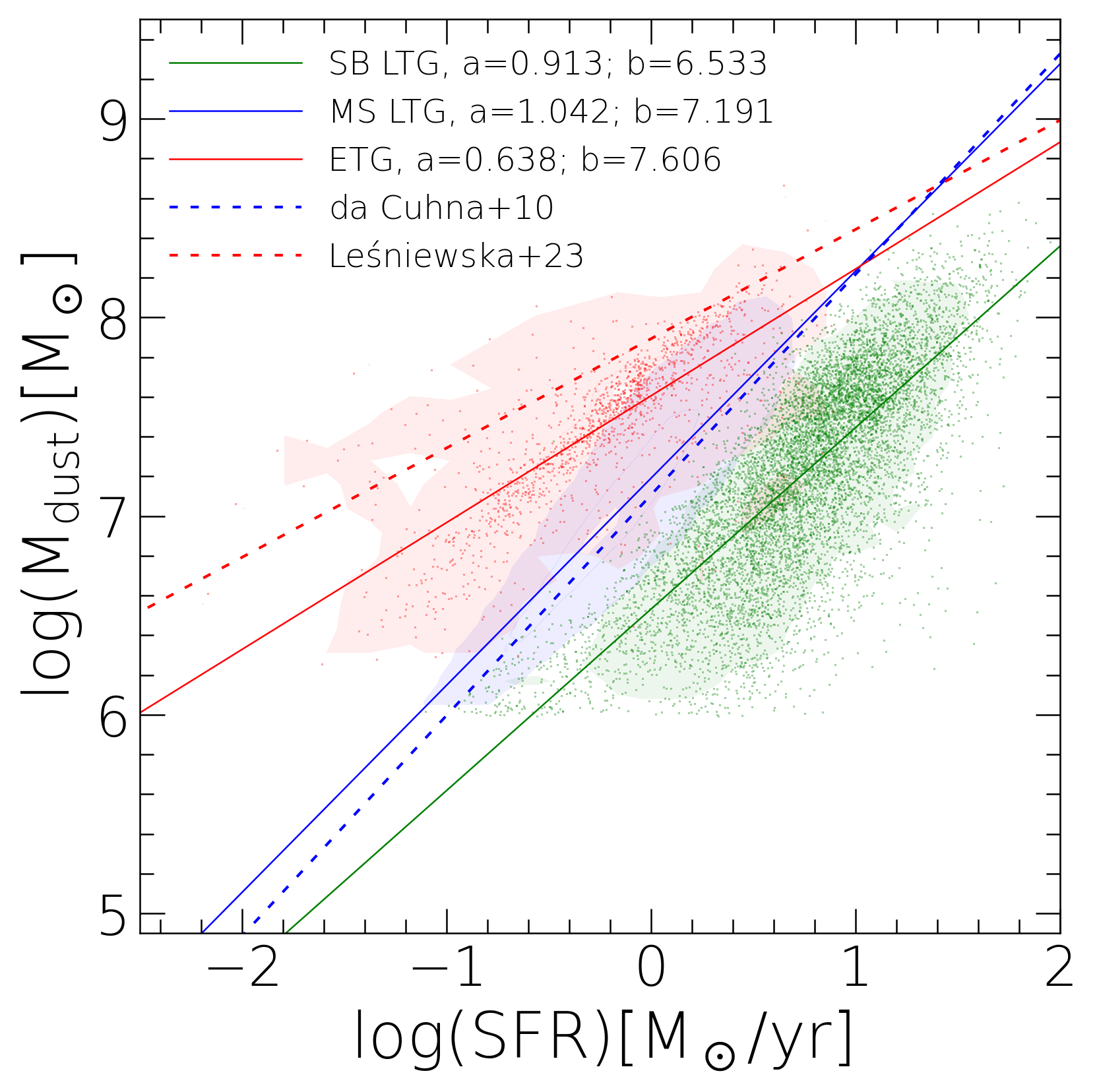}
\caption{Dust mass as a function of the SFR. 
The colours of the contours, lines, and points correspond to: ETGs (red),
MS LTGs (blue), and SBGs (green).  
Power-law
fits are described as $\log(M_{\rm dust}) = a \times \log(SFR) + b$ with
coefficients given in the legend. Red and blue dotted lines show fits from
\citet{Lesniewska2022} and \citet{daCuhna2010}, respectively. Individual
MS LTG are not shown for the sake of clarity.
\label{fig3:daCuhna}}
\end{figure}

In Fig. \ref{fig3:daCuhna}, we show the simulated galaxies and compare them
to the observational results \citep{daCuhna2010,Lesniewska2022}. To this
end, we applied a specific sample selection of starburst galaxies (SBGs), 
normal SFGs, and quenched ETGs below the MS. The majority of the galaxies
studied in \citet{daCuhna2010} are normal MS galaxies with late-type morphologies
in a similar redshift range to the one used in this work. Thus, we selected
galaxies with bulge-to-total mass ratio below 0.3, above the dust mass limit
(see the right panel in Fig. \ref{fig1:MS_dustLim}) and with the same redshift
limit as described in Sect. \ref{sec:data_simulations_lgalaxies}. These constitute
the MS LTG sample. In the context of this particular relation, \citet{Lesniewska2022}
studied ETGs divided between galaxies on the MS and below the MS. Thus,
in addition to the already described criteria (Sect. \ref{sec:data_simulations_lgalaxies}),
 we also selected galaxies that are below -0.2 dex from the simulated MS
(Fig. \ref{fig1:MS_dustLim}, left panel). While this MS has a steeper slope
than the one used in \citet{Lesniewska2022}, we can see that both are consistent
in the stellar mass range considered in this work. In this way, we selected
68\% of the ETG sample below the MS and  used it in this exercise. 

The relation between dust mass and SFR for MS LTG  agrees with the observed
relation from \citet{daCuhna2010} and is described as:
\[\log(M_{\rm dust, MS\,LTG})= 1.042\times \log({\rm SFR}) + 7.191,\] 
with a $2\sigma$ scatter around the fit of 0.24\,dex.

In the case of ETGs, the slope of the fit to simulated galaxies (0.638) is
similar to the one from the observed galaxies \citep[0.55;][]{Lesniewska2022},
but with a $\sim0.3\,$dex of difference in the normalisation parameter, namely,
our ETG sample below the MS have a lower dust content than the observed one.
The relation is as follows:
\[\log(M_{\rm dust, ETG})= 0.638\times \log({\rm SFR}) + 7.606,\] 
with a $2\sigma$ scatter of 0.23\,dex.

In Fig. \ref{fig3:daCuhna}, we also included the relation of dust mass and
SFR for simulated SBGs (green contours and line). These were selected (i)
as galaxies that are five times above the simulated MS, (ii) with a dust
mass is above the dust mass limit estimated from observations (Fig. \ref{fig1:MS_dustLim},
right panel), and (iii) in the same redshift range (Sect. \ref{sec:data_simulations_lgalaxies}).
These objects form a sequence that runs almost parallel with the MS galaxies,
described as:
\[\log(M_{\rm dust, SB})= 0.913\times \log({\rm SFR}) + 6.533,\]
with $2\sigma$ scatter of 0.32\,dex. 

Additionally, we tested  whether the relations given above hold upon the
selection based on star formation activity. To do this, we selected active
(sSFR > 1/t$_{\rm H}$), and passive (sSFR < 1/t$_{\rm H}$) galaxies without
any selection regarding morphology (i.e. bulge-to-total mass ratio), but
maintaining the dust mass limit, stellar mass, and redshift range studied
in this work as well as in \citet{daCuhna2010} and \citet{Lesniewska2022}.
We find excellent agreement with both relations, between the MS LTG and active
samples as well as between the ETG and passive samples (see Fig. \ref{fig_apx:daCuhnaRel}).

\begin{figure*}[t!]
\includegraphics[width=0.41\textwidth,clip]{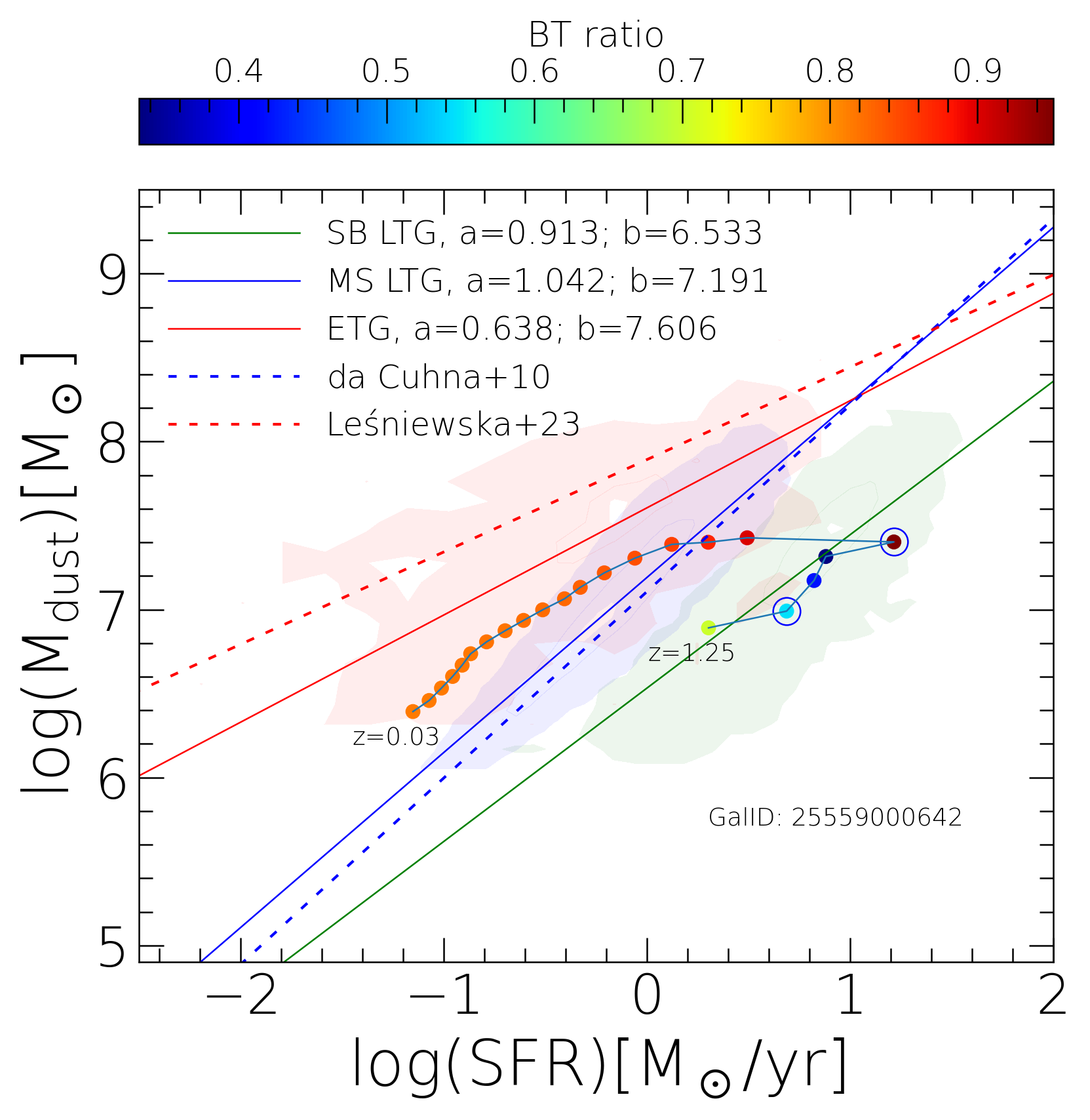}
\includegraphics[width=0.41\textwidth,clip]{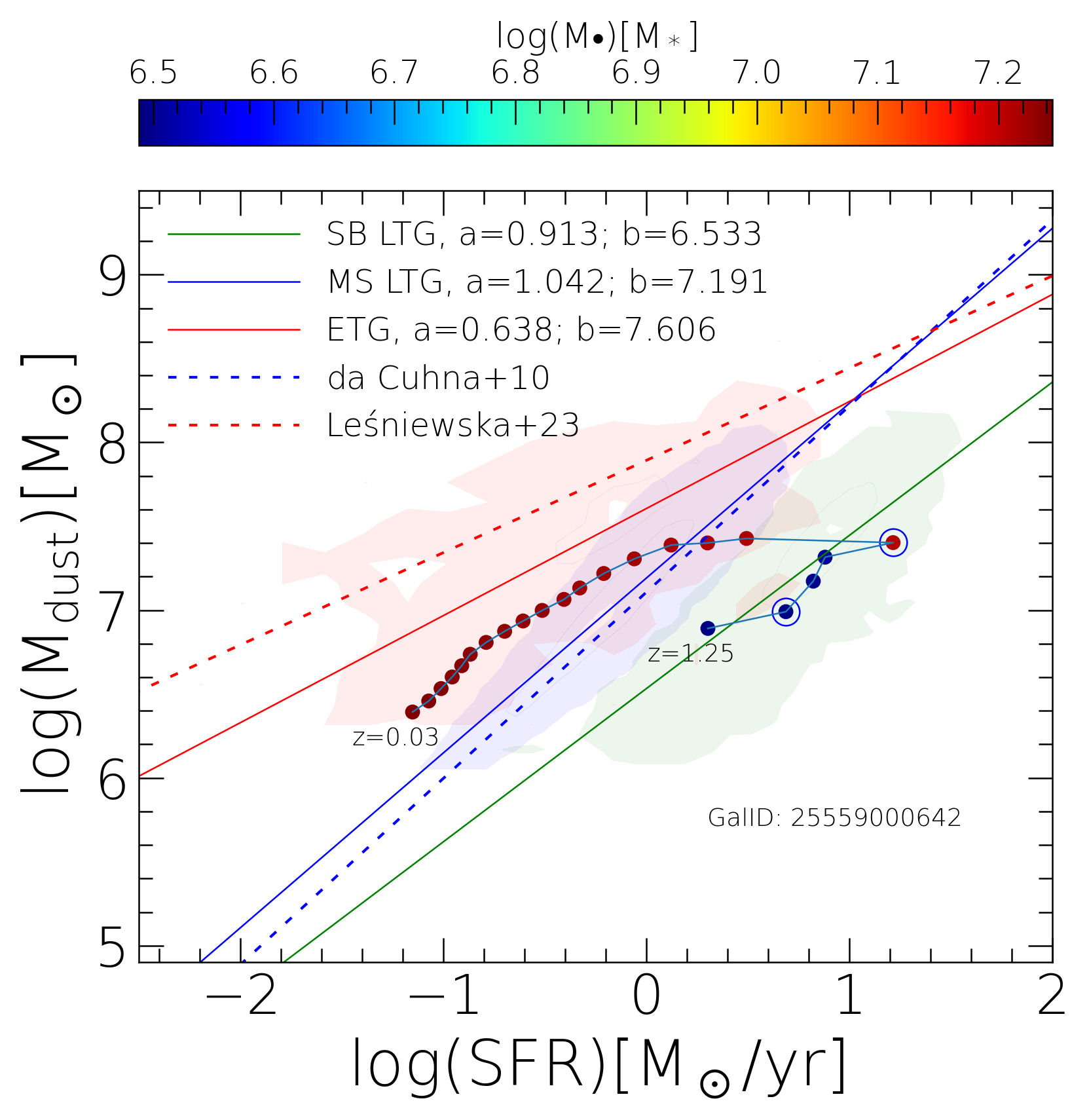}

\includegraphics[width=0.41\textwidth,clip]{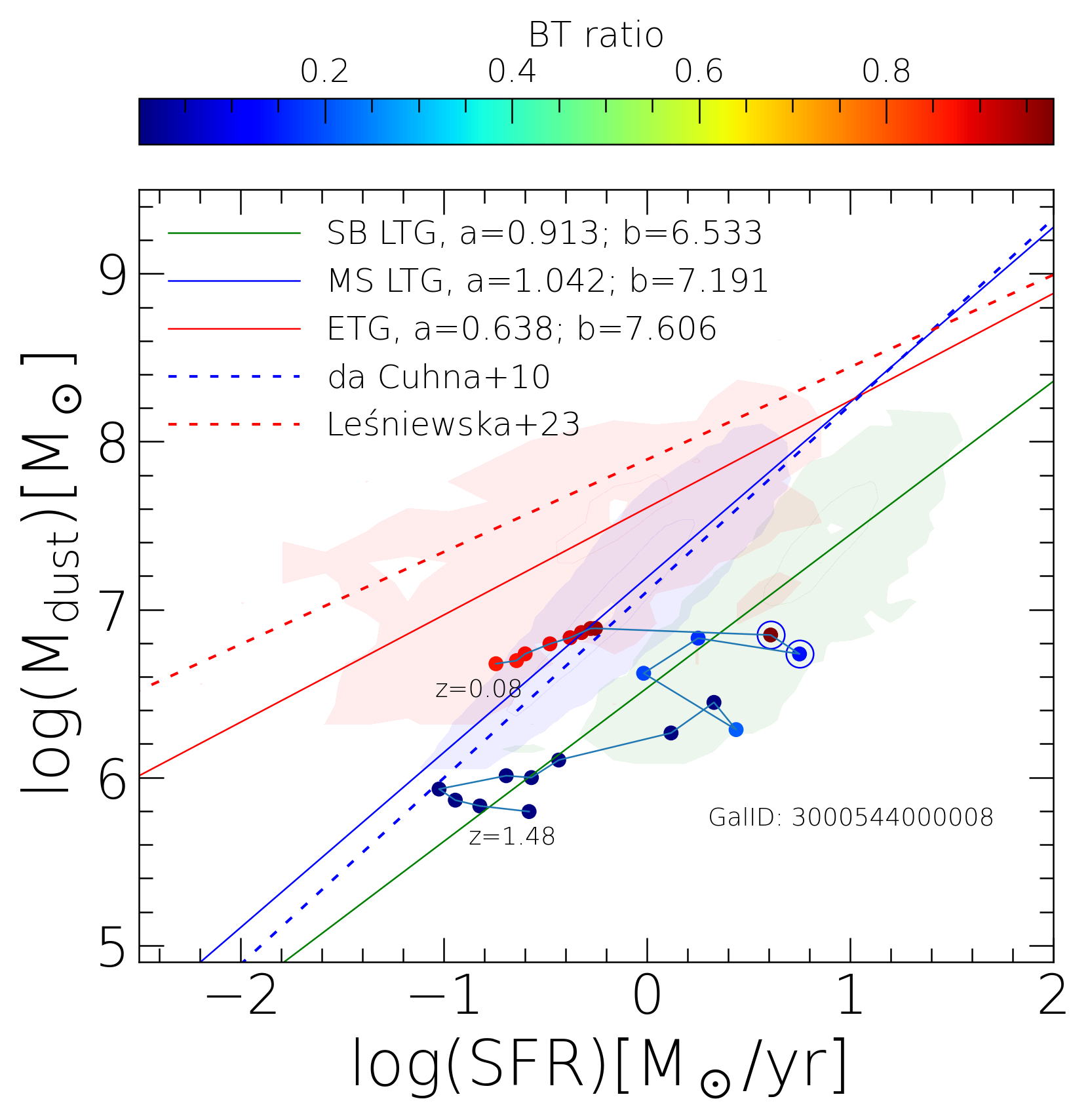}
\includegraphics[width=0.41\textwidth,clip]{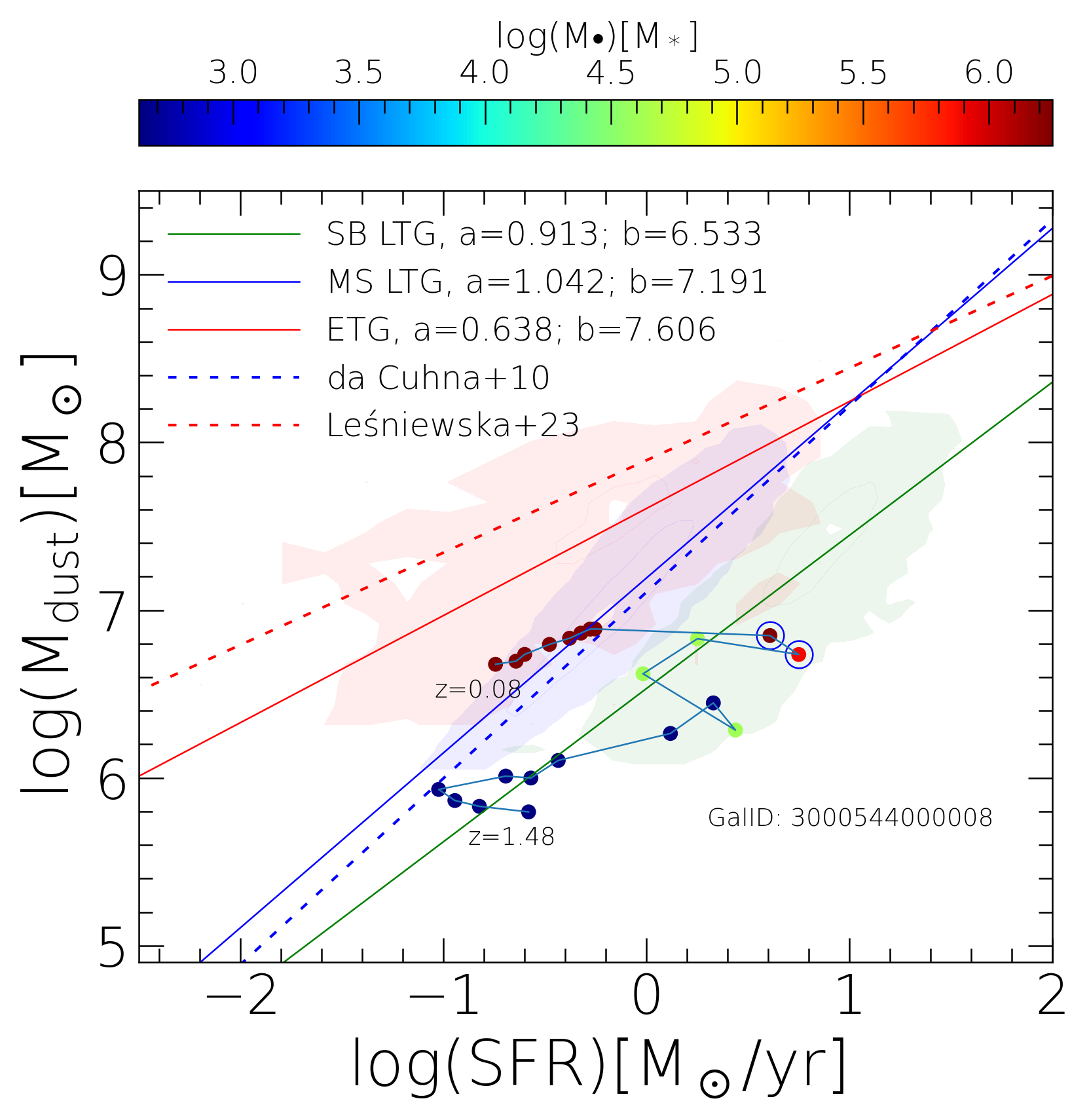}

\includegraphics[width=0.41\textwidth,clip]{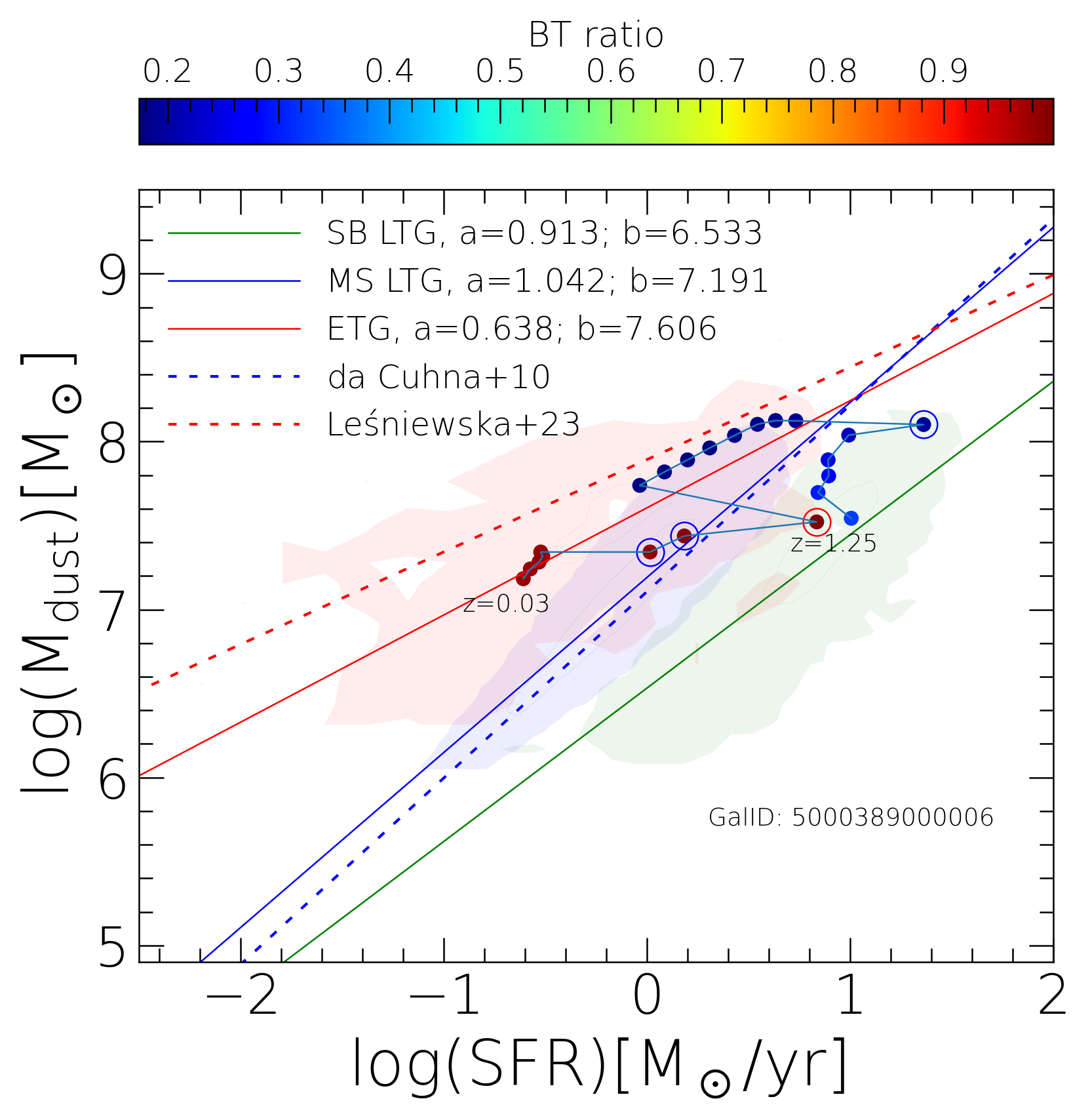}
\includegraphics[width=0.41\textwidth,clip]{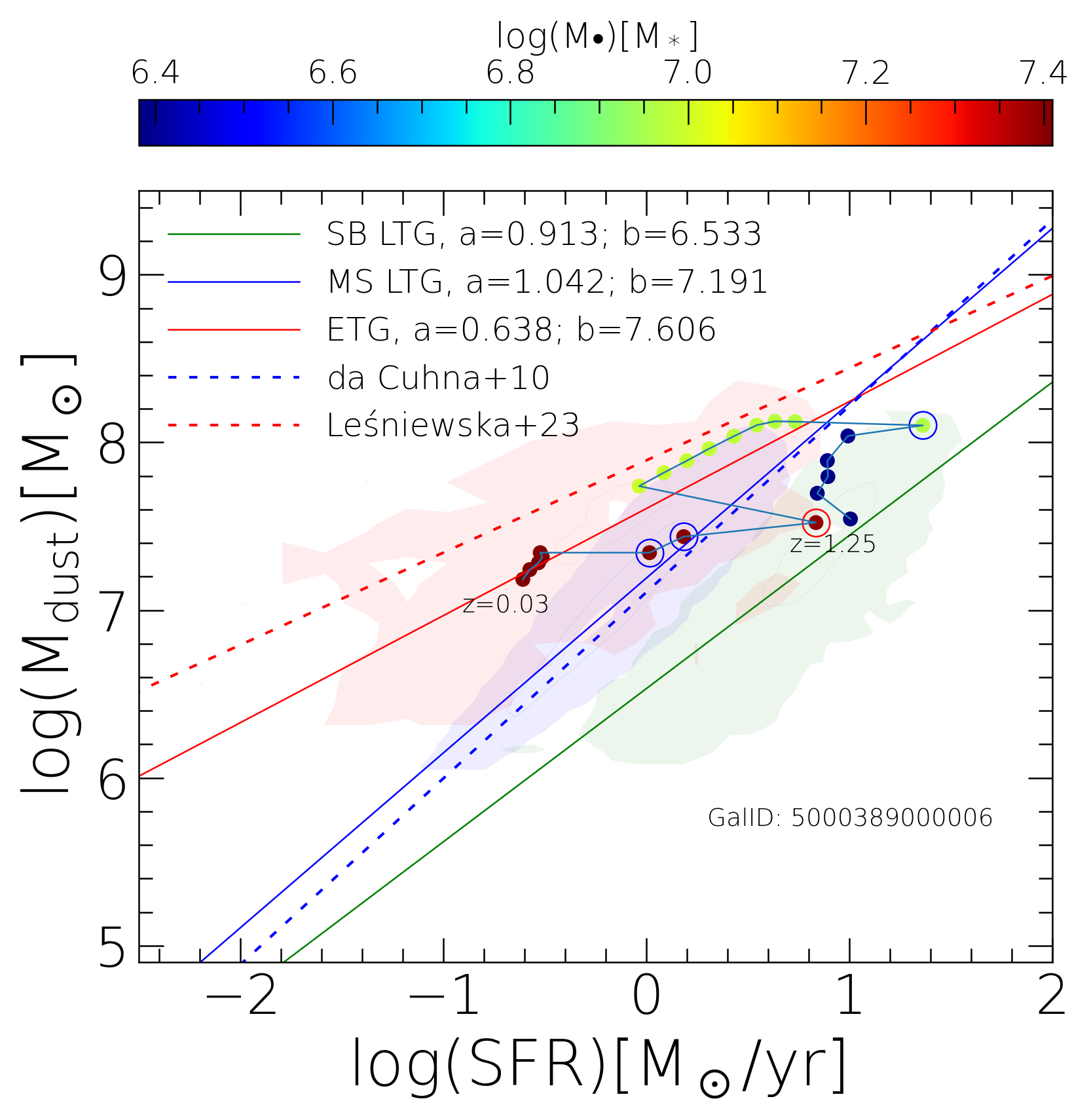}
\caption{Dust mass as a function of the SFR. Progenitors of the selected
ETGs 
are connected with the lines. Empty red (blue) circles show the progenitor
galaxy with a major (minor) merger event 
before that
snapshot. Each row corresponds to one selected ETGs. The left column shows
the colour-coded bulge-to-total mass ratio, whereas the right shows the same
galaxy but  colour-coded by the black hole mass.
\label{fig_apx:dust_sfr}}
\end{figure*}

Figure \ref{fig_apx:dust_sfr} shows the same relation for selected ETGs
and their progenitors; namely, we trace the evolution of a particular galaxy
backward in time. 
This allows us to identify any merger event together with the redshift and
stellar masses of galaxies involved in that event. We define major (minor)
mergers as those with a stellar mass ratio of the merging galaxies less (more)
than three. Tracing merger events, we find that 92\% of our sample went through
at least one (four, on average) minor merger event, while 39\% had at least
one (two, on average) major merger event since $z\sim\,2$. Only about 3\%
of our ETGs sample had no merger events (see Fig. \ref{fig_apx:dust_sfr_no_mergers}).
To put our ETG sample into the broader context of galaxy evolution, we estimated
the major and minor merger events for SBGs and MS LTGs. We find that 75\%
(66\%) of SBGs (MS LTG) had minor mergers, 11\% ($2\%$) had major mergers,
and 22\% (38\%) had no merger events. 
Hence,
it is more likely to select an ETG that had a major or minor merger at some
point in its evolution than a MS LTG or SBGs; however, it is less likely
to find an ETG without any merger than to find a SFG with no merger events
in the past. Starburst galaxies are in between the ETG and MS LTG samples
regarding these fractions. 

The overall picture found in this work is consistent with the models and
interpretation from \citet{Hjorth2014ApJ...782L..23H}, summarised as follows.
Galaxies began to increase their dust mass with a relatively constant (and
high) SFR. The merger events alter the SFR, increasing it for a brief time,
followed by an equally sharp decrease. Finally, a morphologically changed
galaxy (higher bulge-to-total mass ratio) continues its evolution with decreasing
SFR and dust mass in the area where quenched (or passive) galaxies are found.
These objects also have the most massive black holes (see Fig. \ref{fig_apx:sfr_bhmass}).
The morphology of galaxies without any merger event is shaped solely by 
disc instabilities, namely, a mechanism that is efficient enough to drive
morphological transformation, as shown in \citet[][at least in intermediate-mass
galaxies]{Parente2023MNRAS.tmp..881P}.

\subsection{Dust removal mechanisms}\label{sec:dust_removal_mechanisms}

\begin{figure}[t!]
\includegraphics[width=0.48\textwidth,clip]{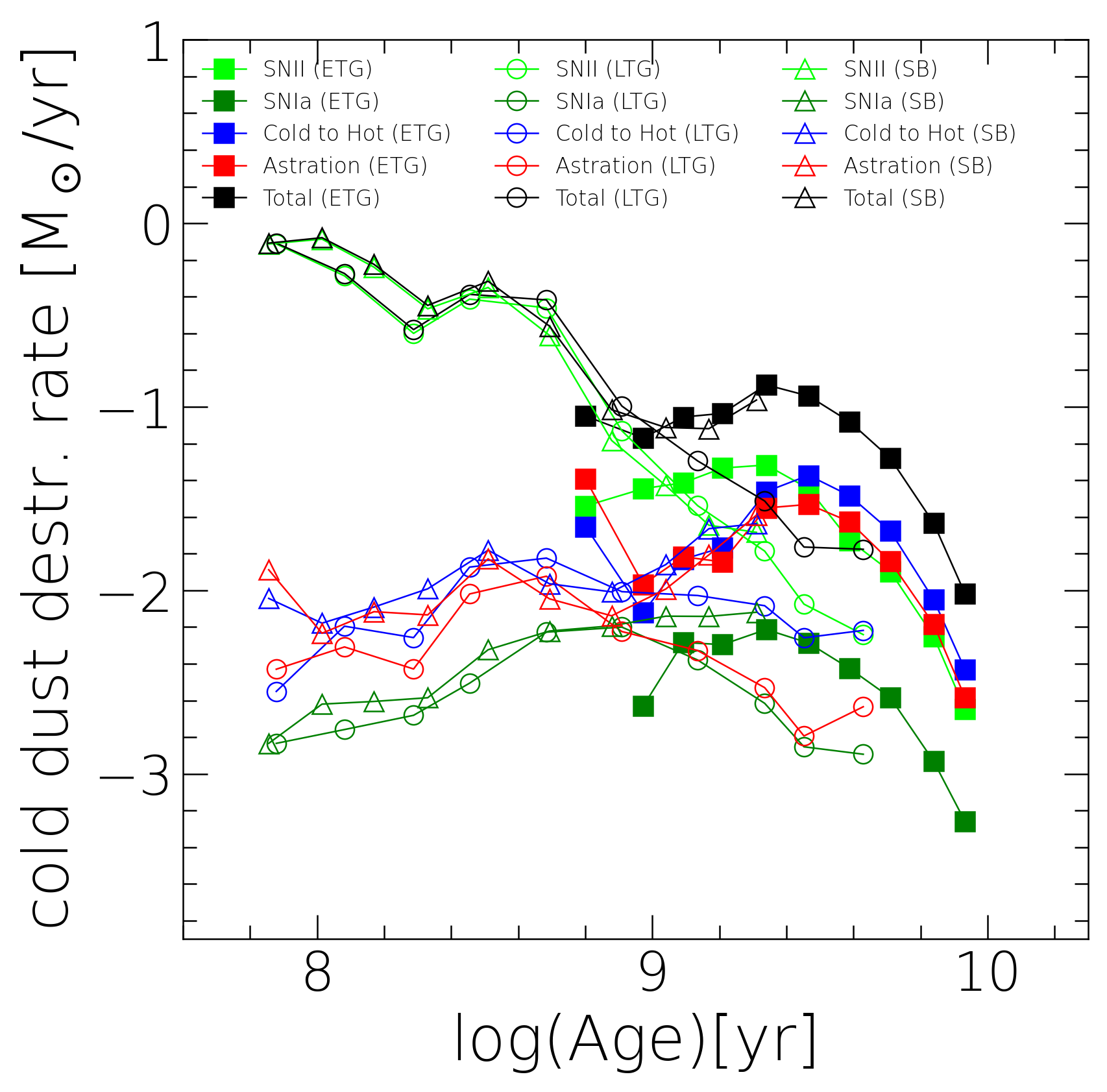}
\caption{Dust destruction rates as a function of light-weighted age. ETGs,
MS LTGs, and SBGs are shown as filled squares, empty circles, and triangles,
respectively. Colours denote different destruction and removal mechanisms
(see main text for details). Only median values in each age bin are shown
for the sake of clarity. 
\label{fig7:dust_destr_rates}}
\end{figure}

Taking advantage of the simulations, it is possible to find the dominant
mechanism of dust removal. 
In particular, we analysed the dust destruction (or removal) rate by SNII
and SNIa (SN shocks), the rate of the transition from the cold to hot phase
(i.e. ejected from the cold gas as a consequence of the heating from stellar
feedback), and astration (the use of cold dust for star formation). These
destruction rates are shown in Fig. \ref{fig7:dust_destr_rates} for our ETG
sample, as well as for LTGs and SBGs (for reference). For LTGs and SBGs,
the general trends are similar: the shocks from SNII are dominating (by 2--3
dex) over the rest of the mechanisms for the majority of the age span. For
 ETGs SNII dominate between log(age/yr)$\sim8.8$ and $9.4$. After that, stellar
feedback dominates, with astration, and SN feedback on a similar level, all
decreasing with age. The age range of the change in the main dust destruction and removal
mechanism is also aligned with the age at which the feedback from AGN radio
mode accretion is strong enough to prevent inflows.
Thus, after heating the gas with stellar feedback, the AGN feedback effectively
prevents its cooling (see Fig. \ref{fig:dust-to-stellar_with_BHmass}), thereby
indirectly decreasing the SFR (and, hence, the SNII rate). {Similar result was found in \citet{Lorenzon2024arXiv240410568L} but using hydrodynamical simulations.}

To test whether some of the destruction mechanism shapes the fitted removal
timescale, we switched off each of these at a time. We find that although
the normalisation changes by about 0.4 dex, the resulting removal timescales
agree with the errors (as shown in Fig. \ref{fig_apx:switchingoff}). Our
results are consistent with \citet[][in particular, as seen in their Fig.
12]{Parente2023MNRAS.tmp..881P}

\section{Discussion and conclusions}
\label{sec:conslusions} 
In this work, we adopted the \lgalaxies\ SAM \citep{HenriquesLGal2020,Parente2023MNRAS.tmp..881P}
to study the ISM removal timescales in ETGs and compare them with recent
observations. We also investigated the relationship between dust mass and
SFR. Furthermore, we also investigated the mechanisms that are responsible
for the removal or destruction of the cold dust content in simulated galaxies. 
The selection process involved the observed dust mass limit, thus the results
may not be applicable to less dusty ETGs.

Using our selected sample, we recovered the observed ISM (dust and \hi) removal
timescales (Fig. \ref{fig1:MS_dustLim}). Our dust and \hi\ removal timescales
of \tauDust=\jntauDust\ and \tauGas=\jntauHI\ Gyr are in excellent agreement
with \tauDust=\altauDust\ and \tauGas=\mmtauHI\ Gyr estimated for our reference
GAMA sample, studied in \citet{Lesniewska2022}. We also refer to a similar
smaller sample in \citet{Michalowski2022}.This confirms that timing the
ISM removal using the stellar age is appropriate for these galaxies (as in
\citealt{Michalowski_dust2019,Michalowski2022} and \citealt{Lesniewska2022}).

We show that the dust removal timescale depends slightly on stellar mass
and BH mass (albeit at a low statistical significance), with almost identical
normalisations with respect to the observationally measured ones. No dependence
of the ISM removal on redshift has been found. 
We do find, however, that using sub-samples divided into stellar mass bins
have different normalisation constants (Eq. \ref{eq:expo}) in the case of
the \hi-to-stellar mass ratio. This is due to the gas-to-dust mass ratio
and its correlation with stellar mass (metallicity), as shown in 
\citet[][in particular, their Fig. 3]{RemyRuyer2014A&A...563A..31R}. This is also true for BH mass bins, however, it is less visible in the redshift bins. 

In the SAM, we used the black hole activity, in particular, the SMBH accretion
of hot gas and the consequent radio mode feedback controls the inflow (cooling)
of the material. There are 95\% (5\%) of galaxies with (without) inflow
in the MS LTG and SBG samples since the last snapshot. In contrast, we find
that 82\% of ETGs had no inflow between the current and the previous snapshot.
The overall dust and \hi\ removal timescales do not change if we exclude
galaxies with inflow (top panels in Fig. \ref{fig:dust-to-stellar_with_BHmass}).
The hot-to-cold gas mass ratio is controlled by the black hole activity,
increasing with increasing black hole mass (bottom left panel in Fig. \ref{fig:dust-to-stellar_with_BHmass}).
Galaxies with higher hot gas masses are also those with low sSFR (or below
the MS; see the bottom right panel in Fig. \ref{fig:dust-to-stellar_with_BHmass}). 
The mechanism that prevents the inflow (radio mode feedback) also retains
a galaxy in a passive state, with low SFR, below the MS (as is the case of
most of the ETGs). On the other hand, galaxies with inflow are galaxies on
and above the MS (as is the case of MS LTGs and SBGs).

Individually, between 63\% and 47\% of our ETGs and their progenitors follow
the dust- and \hi-to-stellar mass ratio evolution with time in the same way
as the fit for the total sample to the dust- and \hi-to-stellar mass ratio
with stellar age. The remaining galaxies and their progenitors have increased
variations in the dust- and \hi-to-stellar mass ratio, which increases the
error of the fitting (Fig. \ref{fig2:exponentialFits_individual}).

We show that simulated galaxies follow the relation between dust mass and
SFR for SFGs on the MS \citep{daCuhna2010} and for passive ETGs \citep{Hjorth2014ApJ...782L..23H,Lesniewska2022};
the latter exhibit a lower normalisation constant (by about 0.35 dex).
 To test whether this is due to the excessive dust destruction we investigated
the dust-to-gas (DTG) ratio of our sample. We find that the DTG ratio of
our ETGs follows the relation between DTG and metallicity for the simulated
SFGs \citep[][their Fig. 8]{Parente2023MNRAS.tmp..881P}{ occupying the high
DTG and metallicity end of this relation. This indicates that the difference
in the normalisation constant} is not due to excessive dust destruction,
but it is most likely the reflection of the slightly lower stellar and dust
masses and SFRs of the selected ETG sample, as compared to the ETGs from
\citet[][which is explained in Sect. \ref{sec:data_simulations_lgalaxies}]{Lesniewska2022}.
We also find, for the first time, a sequence for simulated starburst galaxies,
which runs almost parallel to the star-forming galaxies (Fig. \ref{fig3:daCuhna}).

We find that merger events are responsible for the morphological transformation
of simulated ETGs. This leads to a bulge mass increase and black hole growth
followed by an increase of the radio mode accretion. This accretion mode
results in strong feedback which is capable of preventing the hot gas from
cooling. Using the relation between dust mass and SFR, we show that a merger
event changes dramatically the ISM content, and ignites high star formation.
This event is normally followed by steady, and passive evolution of a high
bulge-to-total mass ratio galaxy in the regime of passive ETGs (Fig \ref{fig_apx:dust_sfr}).

Only about 3\% of the ETG sample had no mergers in the past. Instead, disc
instabilities have played the main role in the morphological transformation
of these objects. In Fig. \ref{fig_apx:dust_sfr_no_mergers}, we show examples
of these ETGs and their progenitors. On the other hand, a total of 22\% (38\%)
of SBGs (MS LTGs) had no merger events. This is a considerably higher rate
compared to ETGs. This reflects an earlier evolutionary stage, where disc
instabilities shape the morphology before the merger-driven morphological
transformation occurs.

We investigated different mechanisms of dust destruction implemented in the
adopted SAM. Shocks associated with SN explosions dominate the destruction
rate over a majority of the galaxy evolution in the MS LTGs, SBGs, and younger
ETGs with log(age/yr)$<$9.4. For ETGs with log(age)$>$9.4,
AGN feedback is strong enough to prevent any inflow from the hot gas atmosphere
(no cooling), thus slowing down star formation. Then, stellar feedback begins
to dominate the dust destruction, while astration and destruction by SNII
shock waves are on a similar level (see Fig. \ref{fig7:dust_destr_rates}).

We conclude that SNII-driven destruction dominates in the early stages of
ETG evolution. At later stages, the stellar feedback and astration became
similarly important, while the AGN feedback prevents the inflow of cool gas,
decreasing SFRs. In this scenario, the morphological transformation (mostly
merger-driven) is important for efficient BH growth and an increase of AGN
feedback.


\begin{acknowledgements}
J.N. would like to thank R. Yates for his useful comments regarding the \lgalaxies,
and K. Lisiecki for helpful discussions regarding SIMBA simulations. 
J.N., M.J.M., and A.L.~acknowledge the support of 
the National Science Centre, Poland through the SONATA BIS grant 2018/30/E/ST9/00208.
This research was funded in whole or in part by National Science Centre,
Poland (grant number: 2021/41/N/ST9/02662).  M.J.M.~acknowledges the support
of
the Polish National Agency for Academic Exchange Bekker grant BPN/BEK/2022/1/00110
and the Polish-U.S. Fulbright Commission. AL and JH were supported by a VILLUM
FONDEN Investigator grant (project number 16599). This work was supported
by a research grant (VIL54489) from VILLUM FONDEN.
For the purpose of Open Access, the author has applied a CC-BY public copyright
licence to any Author Accepted Manuscript (AAM) version arising from this
submission.
The Millennium Simulation databases used in this paper and the web application
providing online access to them were constructed as part of the activities
of the German Astrophysical Virtual Observatory (GAVO). Authors acknowledge
the use of astropy libraries \citep{2013A&A...558A..33A,2018AJ....156..123A},
as well as TOPCAT/STILTS software \citep{Taylor2005ASPC..347...29T}.
\end{acknowledgements}

\bibliographystyle{aa} 
\bibliography{biblography} 

\begin{appendix}
\section{Dust destruction and removal mechanisms}

\begin{figure}[t!]
\includegraphics[width=0.46\textwidth,clip]{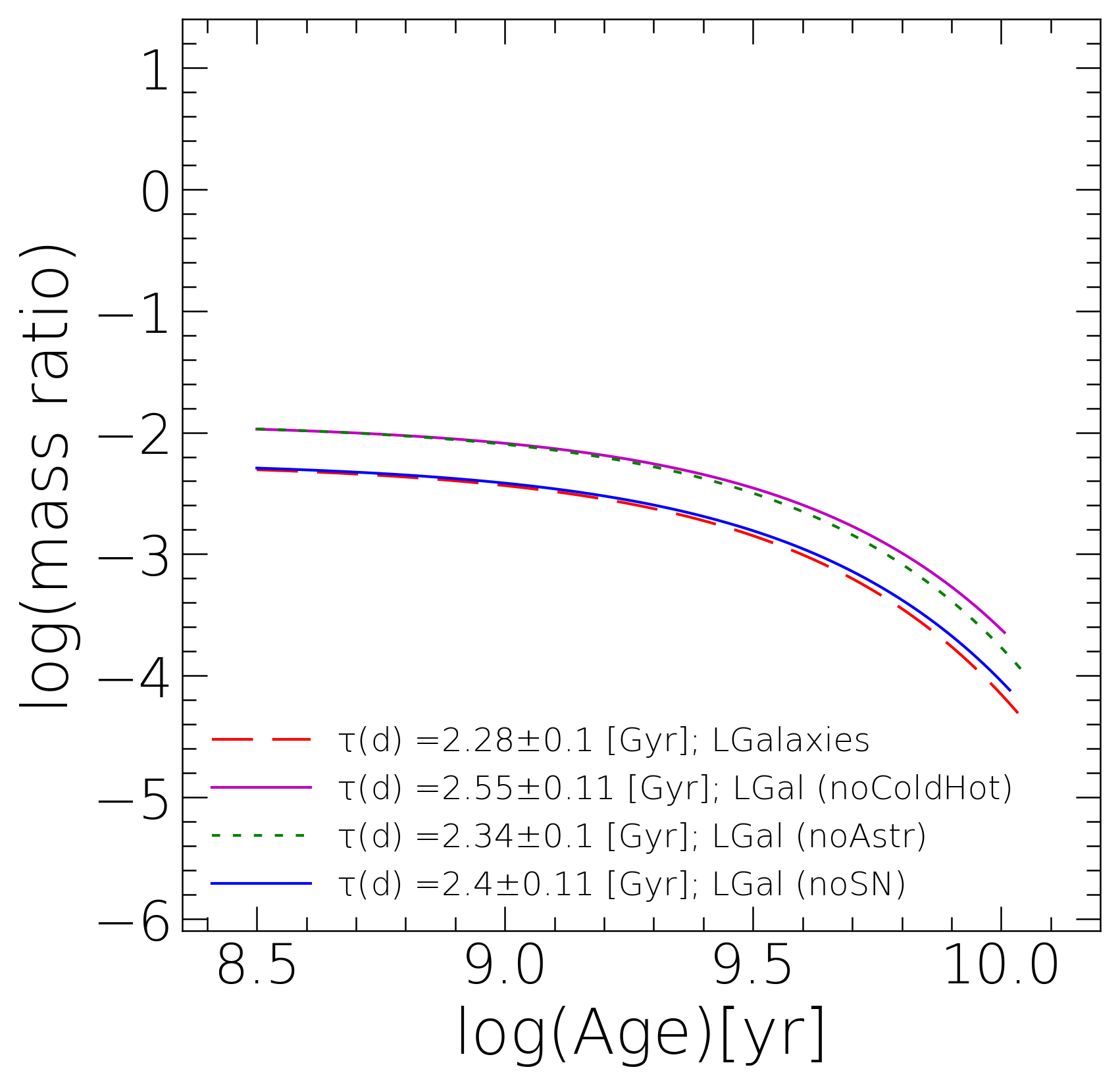}
\caption{Same as Fig. \ref{fig2:exponentialFits}, but with different dust
destruction-and-ejection mechanisms switched off. 
\label{fig_apx:switchingoff}}
\end{figure}

Here, we show the results of the SAM predictions with different dust destruction
and ejection mechanisms switched off. In particular, new runs with no SN
destruction ({\it noSN}), no astration ({\it noAstr}), and no ejection from
cold gas ({\it noColdHot}) were performed. Figure \ref{fig_apx:switchingoff}
shows the resulting exponential fits with removal timescales varying between
2.34 and 2.55 [Gyr], this is within 10\% of the fiducial model. Thus, the
general ISM removal timescale found in used SAM does not depend on the individual
removal and ejection mechanisms. Considering the normalisation of the exponential
fits, we can see that {\it noSN} model gives a very similar dust-to-stellar
mass ratio as the fiducial model. In the fiducial model, the gas phase metals
produced by the SNe destruction are quickly accreted again into grains by
the accretion process. Thus, in the {\it noSN} model this does not happen,
thus accretion is also reduced and the resulting dust content is roughly
similar to the fiducial model. Considering {\it noAstr} and {\it noColdHot}
models, the removal timescale is similar, but the normalisation is higher.
This is because the two processes remove the dust (and metals) from the cold
gas, while the SNe transfer the dust metals into the gas phase, leaving them
available in the cold phase \citep{Parente2023MNRAS.tmp..881P}. 

\section{Dust mass and SFR}
\begin{figure}[t!]
\includegraphics[width=0.46\textwidth,clip]{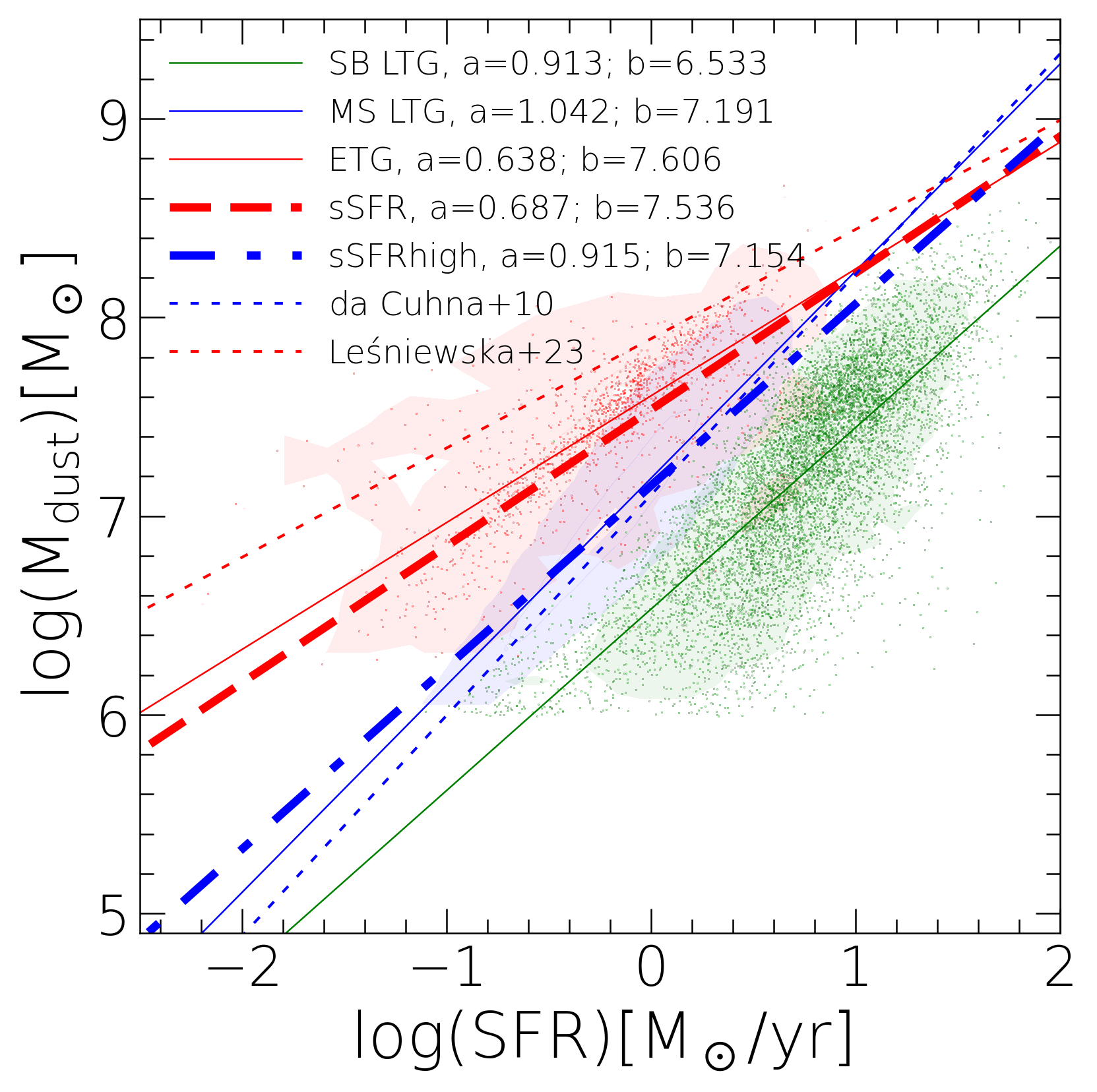}
\caption{Dust mass as a function of SFR as a function of BH mass. Colours
are the same as in Fig. \ref{fig3:daCuhna}, with additional thick blue and
red dashed lines showing a relation for active (sSFR > 1/t$_{\rm H}$) and
passive galaxies respectively (sSFR < 1/t$_{\rm H}$).
\label{fig_apx:daCuhnaRel}}
\end{figure}

To test whether our morphologically based selection indeed corresponds to
a passive galaxy sample, here we provide a selection based on sSFR. We select
active galaxies using sSFR > 1/t$_{\rm H}$, while passive galaxies are defined
as sSFR < 1/t$_{\rm H}$. Additionally, we also keep the dust mass limit,
stellar mass limit and redshift range to be able to make a direct comparison
with observations as well as with the results described in Sect. \ref{sec:dust_and_star_formation}.
As shown in Fig. \ref{fig_apx:daCuhnaRel} the results obtained using sSFR
to select active and passive galaxies are similar to the selection based
on morphology, with $2\sigma$ scatter of 0.23 and 0.26, respectively. The
fitting coefficients are given in the legend and we can see that both the
slopes and intercepts are in agreement.

\section{SFR and black hole mass}
\begin{figure}[t!]
\includegraphics[width=0.46\textwidth,clip]{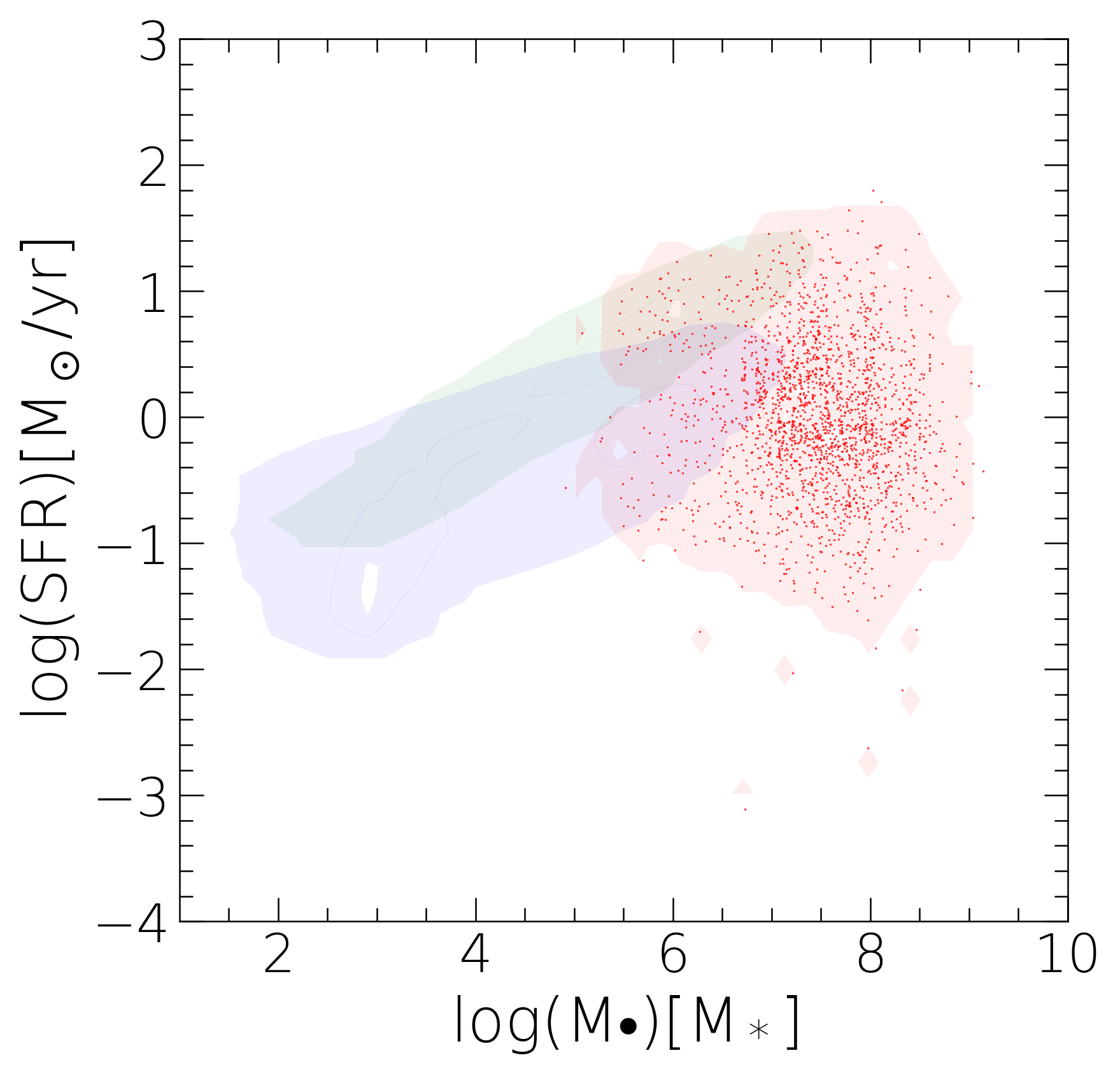}
\caption{SFR as a function of BH mass. In blue, green, and red we show SFGs,
SBGs, and ETGs. 
\label{fig_apx:sfr_bhmass}}
\end{figure}

Figure \ref{fig_apx:sfr_bhmass} shows SFRs as a function of BH masses for
the MS LTGs, SBGs, and ETGs. The most massive BHs are found in simulated
ETGs. The growth of the bulge in a galaxy is likely followed by the BH mass
growth. 

\section{SFR and dust masses for galaxies without a merger event}
\begin{figure*}[t!]
\includegraphics[width=0.41\textwidth,clip]{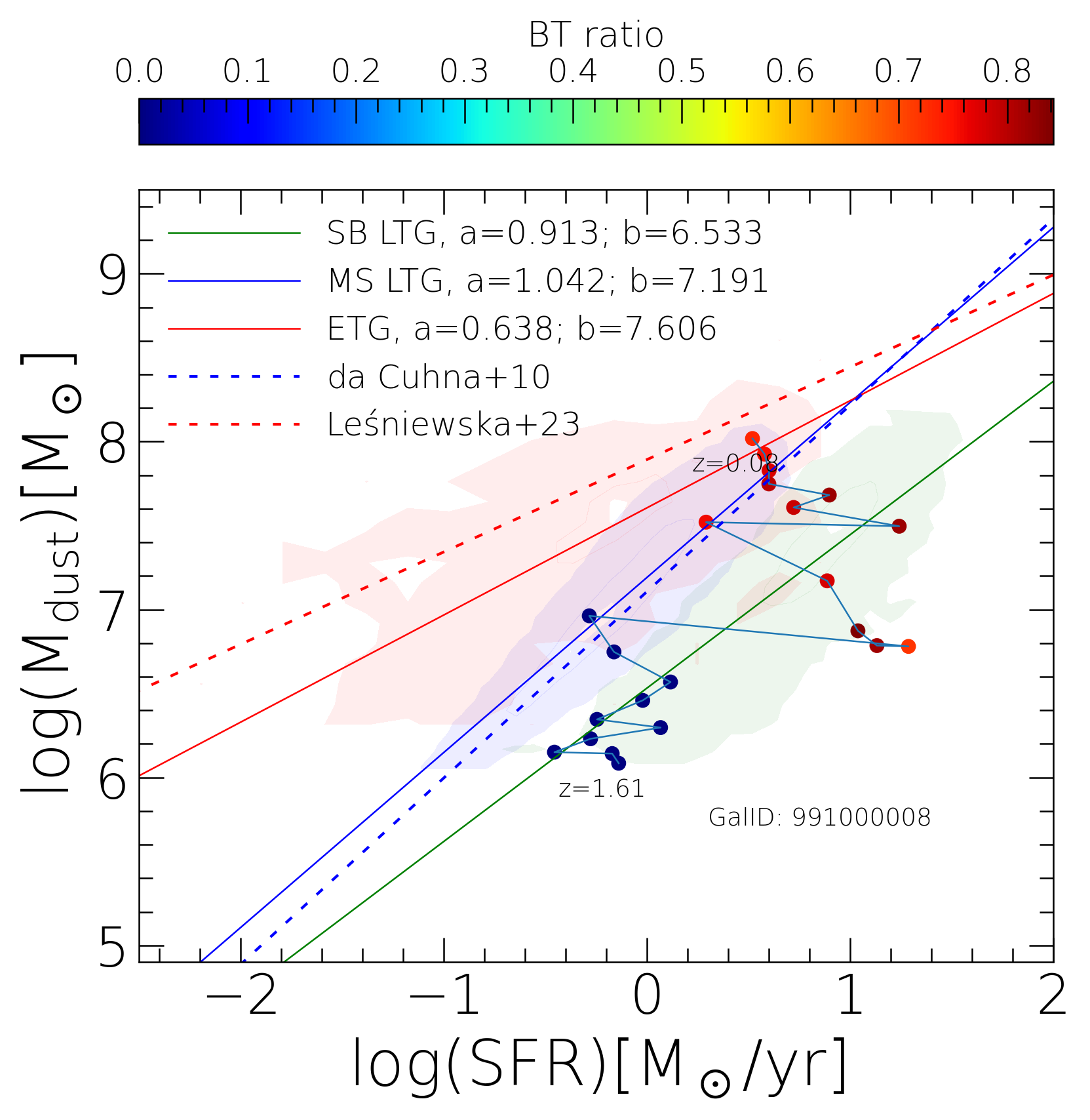}
\includegraphics[width=0.41\textwidth,clip]{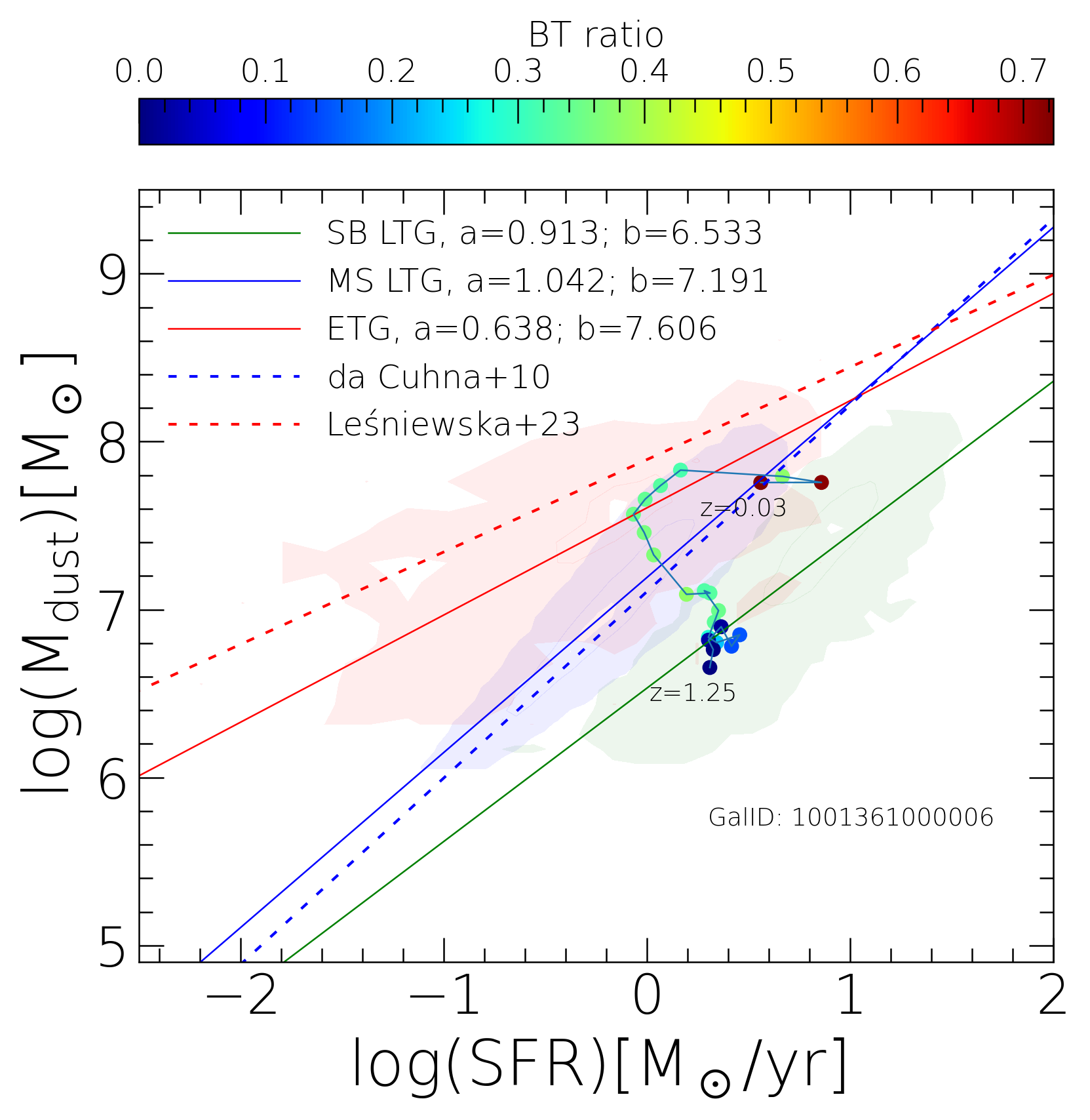}

\includegraphics[width=0.41\textwidth,clip]{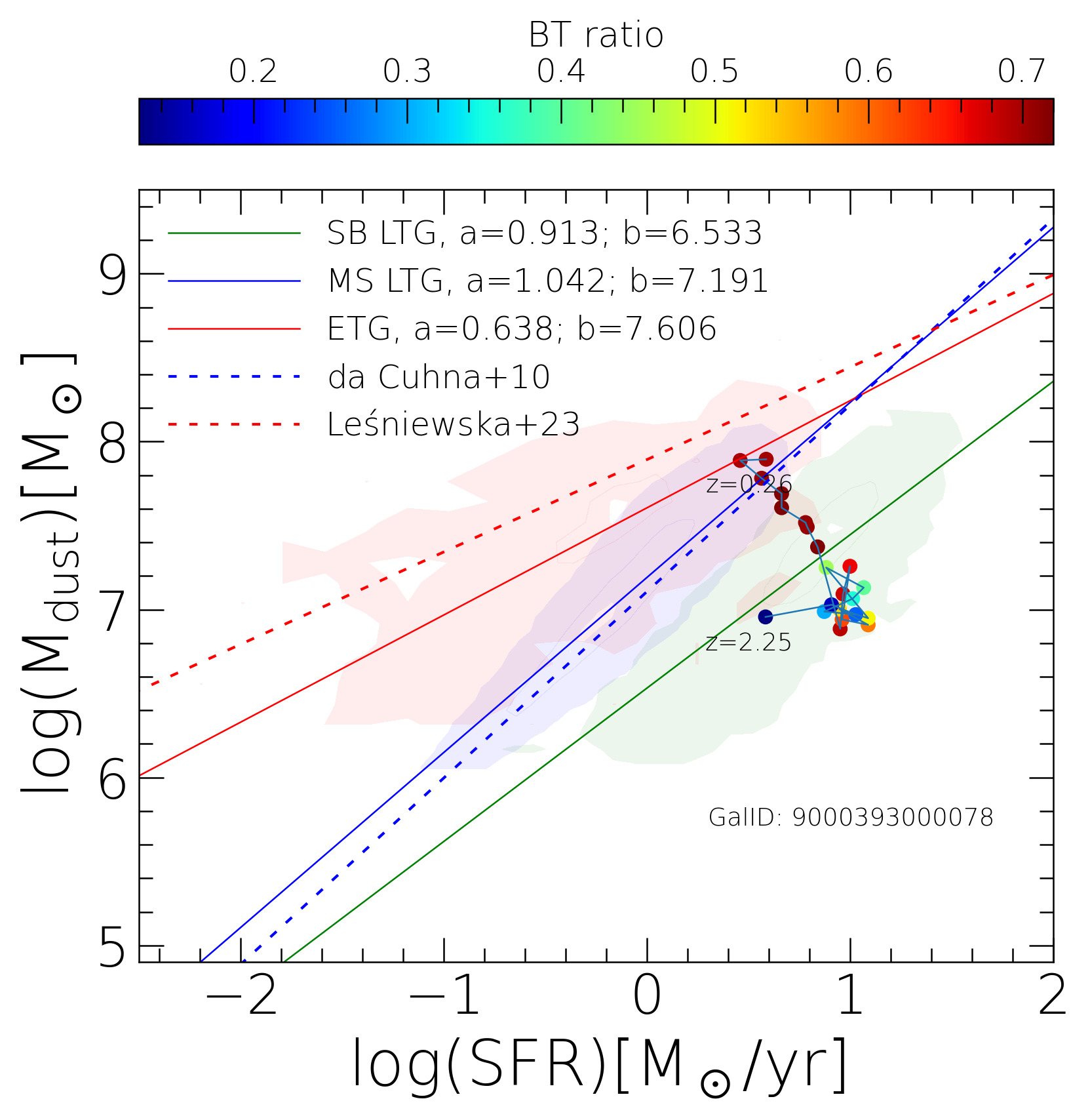}
\includegraphics[width=0.41\textwidth,clip]{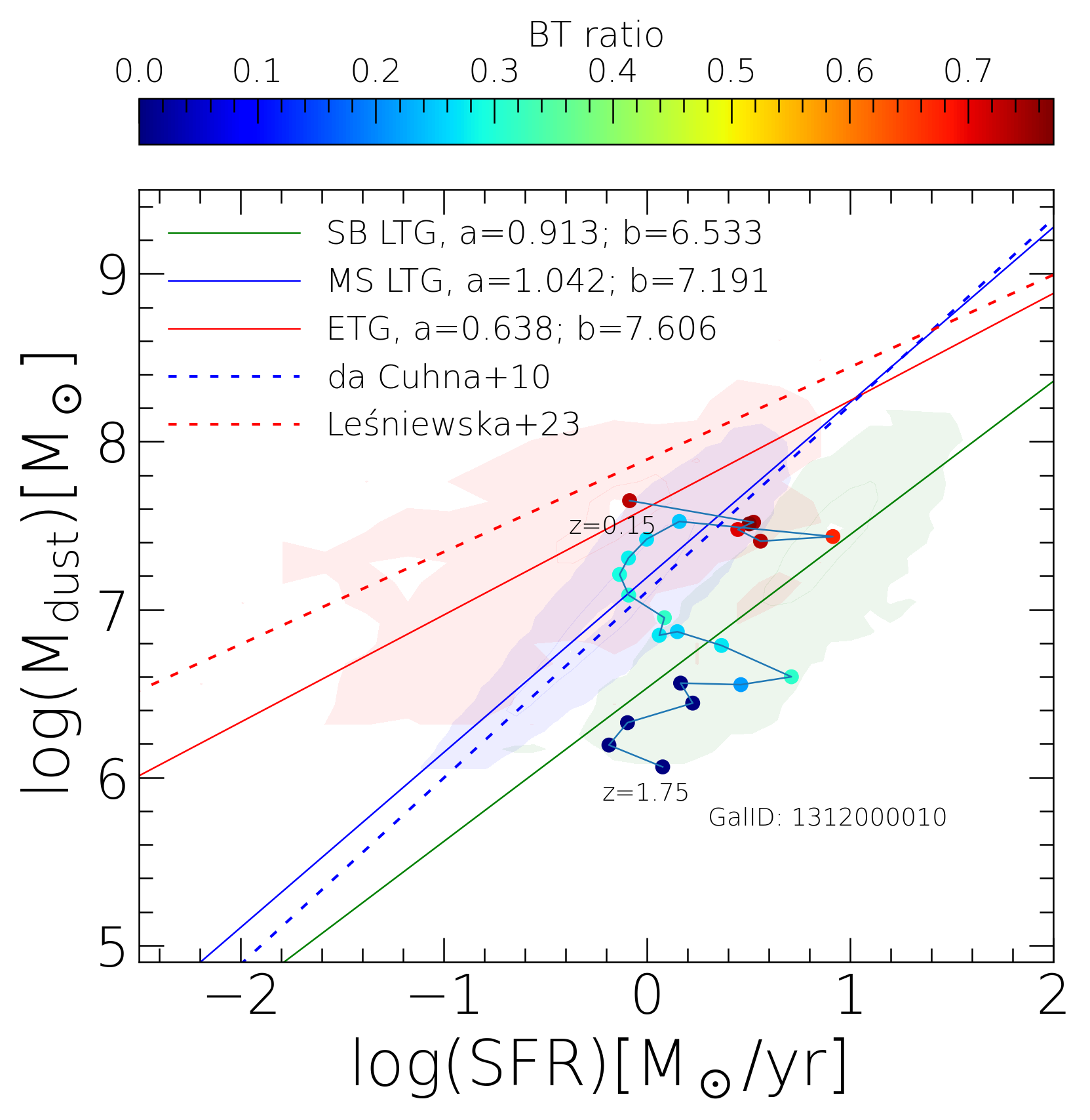}
\caption{ Same as Fig.~\ref{fig_apx:dust_sfr} but for galaxies without a
merger event during their evolution.\  The bulge-to-total mass ratio is colour-coded.
\label{fig_apx:dust_sfr_no_mergers}}
\end{figure*}

In Fig. \ref{fig_apx:dust_sfr_no_mergers}, we show examples of simulated
ETGs from our sample that did not go through a merger event during its evolution.
These constitute $\sim3\%$ of the sample. Their morphological transformation
was driven by disc instabilities as implemented in the SMA used in this work.
For details, see \citet{Parente2023MNRAS.tmp..881P}.

\end{appendix}

\end{document}